\documentclass[12pt]{article}
\usepackage{amsmath}
\usepackage{graphicx,psfrag,epsf}
\usepackage{enumerate}
\usepackage{natbib}
\usepackage{authblk}
\usepackage{url} % not crucial - just used below for the URL
\usepackage[latin9]{inputenc}
\usepackage{float}
\usepackage{mathtools}
\usepackage{amsmath}
\usepackage{amssymb}
\usepackage{setspace}
\usepackage{esint}
\usepackage{appendix}
\usepackage{cleveref}
\makeatletter

\providecommand{\tabularnewline}{\\}
\floatstyle{ruled}
\newfloat{algorithm}{tbp}{loa}
\providecommand{\algorithmname}{Algorithm}
\floatname{algorithm}{\protect\algorithmname}

\usepackage{latexsym}
\usepackage{amsthm}\usepackage{amsfonts}\usepackage{graphicx}\usepackage{epsfig}
\usepackage{bm}
\newcommand*{\patchAmsMathEnvironmentForLineno}[1]{%
      \expandafter\let\csname old#1\expandafter\endcsname\csname #1\endcsname
      \expandafter\let\csname oldend#1\expandafter\endcsname\csname end#1\endcsname
      \renewenvironment{#1}%
         {\linenomath\csname old#1\endcsname}%
         {\csname oldend#1\endcsname\endlinenomath}}%
    \newcommand*{\patchBothAmsMathEnvironmentsForLineno}[1]{%
      \patchAmsMathEnvironmentForLineno{#1}%
      \patchAmsMathEnvironmentForLineno{#1*}}%
    \AtBeginDocument{%
    \patchBothAmsMathEnvironmentsForLineno{equation}%
    \patchBothAmsMathEnvironmentsForLineno{align}%
    \patchBothAmsMathEnvironmentsForLineno{flalign}%
    \patchBothAmsMathEnvironmentsForLineno{alignat}%
    \patchBothAmsMathEnvironmentsForLineno{gather}%
    \patchBothAmsMathEnvironmentsForLineno{multline}%
    }
\usepackage[mathlines,displaymath]{lineno}
\include{def}
\def\dispmuskip{\thinmuskip= 3mu plus 0mu minus 2mu \medmuskip=  4mu plus 2mu minus 2mu \thickmuskip=5mu plus 5mu minus 2mu}
\def\textmuskip{\thinmuskip= 0mu                    \medmuskip=  1mu plus 1mu minus 1mu \thickmuskip=2mu plus 3mu minus 1mu}
\def\beq{\dispmuskip\begin{equation}}    \def\eeq{\end{equation}\textmuskip}
\def\beqn{\dispmuskip\begin{displaymath}}\def\eeqn{\end{displaymath}\textmuskip}
\def\bea{\dispmuskip\begin{eqnarray}}    \def\eea{\end{eqnarray}\textmuskip}
\def\bean{\dispmuskip\begin{eqnarray*}}  \def\eean{\end{eqnarray*}\textmuskip}

\include{def}
\def\dispmuskip{\thinmuskip= 3mu plus 0mu minus 2mu \medmuskip=  4mu plus 2mu minus 2mu \thickmuskip=5mu plus 5mu minus 2mu}
\def\textmuskip{\thinmuskip= 0mu                    \medmuskip=  1mu plus 1mu minus 1mu \thickmuskip=2mu plus 3mu minus 1mu}
\def\beq{\dispmuskip\begin{equation}}    \def\eeq{\end{equation}\textmuskip}
\def\beqn{\dispmuskip\begin{displaymath}}\def\eeqn{\end{displaymath}\textmuskip}
\def\bea{\dispmuskip\begin{eqnarray}}    \def\eea{\end{eqnarray}\textmuskip}
\def\bean{\dispmuskip\begin{eqnarray*}}  \def\eean{\end{eqnarray*}\textmuskip}

\newcommand{\blind}{0}

\newcommand{\dlq}{\lq\lq}
\newcommand{\drq}{\rq\rq}

\usepackage[mathlines,displaymath]{lineno}
\makeatother
\begin{document}
\def\spacingset#1{\renewcommand{\baselinestretch}%
{#1}\small\normalsize} \spacingset{1}

%%%%%%%%%%%%%%%%%%%%%%%%%%%%%%%%%%%%%%%%%%%%%%%%%%%%%%%%%%%%%%%%%%%%%%%%%%%%%%

\if0\blind
{
  \title{Efficient Bayesian inference for multivariate factor stochastic volatility
models with leverage\thanks{The research of Gunawan and Kohn was partially supported
by the ARC Center of Excellence grant CE140100049. The research of all the authors was also partially supported by ARC Discovery Grant DP150104630. }  }
  \author[1]{David G. Gunawan}
    \author[1]{Chris K. Carter}
    \author[1]{Robert J. Kohn}
\affil[1]{School of Economics, University of New South Wales}
  \maketitle
} \fi

\if1\blind
{
  \bigskip
  \bigskip
  \bigskip
  \begin{center}
    {\LARGE\bf Efficient Bayesian inference for multivariate factor stochastic volatility
models with leverage}
\end{center}
  \medskip
} \fi
\bigskip
\begin{abstract}
This paper discusses the efficient Bayesian estimation of a multivariate
factor stochastic volatility (Factor MSV) model with leverage. We
propose a novel approach to construct the sampling schemes that converges
to the posterior distribution of the latent volatilities and
the parameters of interest of the Factor MSV model based on recent advances
in Particle Markov chain Monte Carlo (PMCMC). As opposed to the approach
of \citet{Chib2006} and \citet{Omori2007}, our approach does not
require approximating the joint distribution of outcome and volatility
innovations by a  mixture of bivariate normal distributions.
To sample the free elements of the loading matrix we employ the interweaving
method used in \citet{Kastner:2017} in the Particle Metropolis within
Gibbs (PMwG) step. The proposed method is illustrated empirically
using a simulated dataset and a sample of daily US stock returns.
\end{abstract}

\noindent%
\vfill

%\newpage
\spacingset{1.45} % DON'T change the spacing!

\section{Introduction}

The analysis of financial time series has become an important research
area over the last two decades, with both methodological and computational
developments making  it possible to estimate more complex models.
Two well-known classes of models, the GARCH and stochastic volatility (SV),
have been proposed to model financial time series volatility (see
\citet{Bollerslev:1994} and \citet{Ghysels:1996}). However, current
real world financial applications call for jointly modeling many
simultaneous and co-varying observations over time. Recently, the
literature has dealt with the development of multivariate models
and estimation of such models. Factor multivariate stochastic volatility
(factor MSV) models are increasingly used because they are able to
model the volatility dynamics of a large system of financial or economic
time series when the common features in these series can be captured
by a small number of latent factors. Our article 
focuses on the model formulated by \citet{Chib2006} and extends it to
include leverage.

A computationally efficient method of estimating a high dimensional
factor MSV model is necessary if such models are to be applied to
real world financial applications. Bayesian MCMC methods have been
proposed to estimate the parameters of the factor MSV model (see for
example, \citet{Chib2006,H:2006,Aguilar:2000}). Based on results
reported in the literature, such as \citet{Chib2006}, estimating
a factor MSV using current Bayesian simulation methods is neither exact
nor flexible for two reasons. The first is related
to sampling the latent volatilities. \citet{Chib2006} use the approach
proposed by \citet{Kim:1998} to approximate the joint distribution
of outcome innovations by a suitably constructed seven-component mixture
of normal distributions. The second is related to sampling
the latent factors and the associated free parameters in the loading
matrix. Our aim is to outline a reliable and efficient method
for exact Bayesian inference that performs well and is  easy to implement and extend.

We develop a general approach to constructing sampling schemes that
converge to the correct posterior distribution of the latent volatilities
and the parameters of interest of the Factor MSV based on recent advances
in Particle Markov chain Monte Carlo (PMCMC). The sampling schemes
generate particles as auxiliary variables. \citet{Andrieu:2010} proposed
two particle MCMC samplers. The first is Pseudo Marginal Metropolis-Hastings
(PMMH), where the parameters are generated with the latent states
integrated out. The second is a Particle Gibbs (PG) algorithm. PG
is a Monte carlo approximation of the standard Gibbs sampling procedure
which uses sequential Monte carlo (SMC) to update the states given
the parameters. \citet{Andrieu:2010} shows that the augmented target
density of these two algorithms has the joint posterior density of
the parameters and states as a marginal density. Furthermore, \citet{Mendes2016}
proposed a general PMCMC sampler which combine the PG and PMMH. This
mixed sampler is highly efficient when there is a set of parameters
that is not highly correlated with the latent states which can be
generated using PG, and another set of parameters that is highly correlated
with the latent states and is generated using the PMMH sampler.

In this paper, we develop a version of PG of \citet{Andrieu:2010}
and mixed sampler of \citet{Mendes2016} to sample both the latent
volatilities and the parameters of Factor MSV. Note that in this case,
our approach also does not require to approximate the joint distribution
of outcome and volatility innovations by a ten-component mixture of
bivariate normal distributions \citep{Omori2007}. To sample the free
elements of the loading matrix we employ interweaving method as in
\citet{Kastner:2017} in the Particle Metropolis within Gibbs (PMwG)
step. The proposed method is illustrated empirically using simulated
dataset and a sample of daily US stock returns.

The remainder of this paper is organized as follows. Section \ref{sec:Factor-SV-Model}
describes the Factor MSV model in detail. Section \ref{sec:Proposed-PMCMC-algorithm}
gives an in-depth discussion of the estimation algorithm and its implementation.
Section~\ref{sec:Simulation-Study} presents measures of sampling
efficiency for a simulated dataset. Section \ref{sec:Empirical-Application-to}
discusses an empirical application to US stock returns. Section \ref{sec:Conclusions}
concludes.

\section{Factor SV Model with leverage in the Idiosynchratic Error\label{sec:Factor-SV-Model}}

\subsection{Model\label{SS: model}}

Let $y_{t}=\left(y_{1t},...,y_{pt}\right)^{'}$ denote the $p$ observations
at time $t$ and suppose that conditional on $k$ unobserved factors
$f_{t}=\left(f_{1t},...,f_{kt}\right)^{'}$, we have

\begin{equation}
y_{t}=Bf_{t}+u_{t},\label{eq:standard factor model}
\end{equation}
where $B$ is an unknown $p\times k$ factor loadings matrix of unknown
parameters.
\[
\left(\begin{array}{c}
u_{t}\\
f_{t}
\end{array}\right)\sim N_{p+k}\left(0,\left(\begin{array}{cc}
V_{t} & 0\\
0 & D_{t}
\end{array}\right)\right)
\]
are conditionally independent Gaussian random vectors. The time varying
variance matrices $V_{t}$ and $D_{t}$ are taken to depend upon unobserved
random variables $h_{1t}=\left(h_{11,t},...,h_{1p,t}\right)$ and
$h_{2t}=\left(h_{21,t},...,h_{2k,t}\right)$ in the form
\[
V_{t}=V_{t}\left(h_{1t}\right)=diag\left\{ \exp\left(h_{11,t}\right),...,\exp\left(h_{1p,t}\right)\right\} :p\times p
\]

\[
D_{t}=D_{t}\left(h_{2t}\right)=diag\left\{ \exp\left(h_{21,t}\right),...,\exp\left(h_{2k,t}\right)\right\} :k\times k
\]
where each $h_{1i,t}$ and $h_{2j,t}$ follows an independent three
parameter $\left(\mu_{1i},\phi_{1i},\tau_{1i}^{2},\mu_{2j},\phi_{2j},\tau_{2j}^{2}\right)$
stochastic volatility process
\begin{equation}
h_{1st}-\mu_{1s}=\phi_{1s}\left(h_{1st-1}-\mu_{1s}\right)+\eta_{1st},\eta_{1st}\sim N\left(0,\tau_{1s}^{2}\right),s=1,...,p\label{eq:likelihood SV1}
\end{equation}
and
\begin{equation}
h_{2jt}-\mu_{2j}=\phi_{2j}\left(h_{2jt-1}-\mu_{2j}\right)+\eta_{2jt},\eta_{2jt}\sim N\left(0,\tau_{2j}^{2}\right),j=1,...,k.\label{eq:likelihood SV2}
\end{equation}
We model the joint distributions of outcome innovations and volatilities
as follows

\[
\left(\begin{array}{c}
u_{st}\\
\eta_{1st}
\end{array}\right)\sim N_{p}\left(0,\left(\begin{array}{cc}
\exp\left(h_{1st}\right) & \rho_{1s}\tau_{1s}\exp\left(h_{1st}/2\right)\\
\rho_{1s}\tau_{1s}\exp\left(h_{1st}/2\right) & \tau_{1s}^{2}
\end{array}\right)\right),s=1,...,p,
\]
where $\rho_{s}$ is the correlation coefficient between $u_{st}$
and $\eta_{1st}$ and it is used to measure the leverage effect. \citet{Harvey1996}
were the first to proposed the univariate SV model with leverage effects
in discrete time.

First, to prevent factor rotation and column switching, we follow
the usual convention and set the upper triangular part of $B$ to
zero and $diag\left(B\right)$ non-zero  \citep[e.g.][]{Geweke1996}.
This parameterisation imposes an order dependence. Secondly, the model
is also not identified without identifying the scaling of either the
$k$th column of $B$ or the the variance of $f_{kt}$. The usual
solution is to set the diagonal elements of the factor loading matrix
$B_{jj}$ to one, for $j=1,..,r$, while the level $\mu_{2j,t}$ of
the factor volatilities $h_{2j,t}$ is modeled to be unknown. As noted
by \citet{Kastner:2017}, this approach imposes that the first $k$
variables are leading the factors, and making the variable ordering
dependence stronger. We follow \citet{Kastner:2017} and leave the
diagonal elements $B_{jj}$ unrestricted and set the level $\mu_{2j}$
of the factor volatilities $h_{2j,t}$ to zero for $j=1,...,k$. An
intuitive explanation is that the \dlq leadership\drq of a factor is
shared by several series. Each column of $B$ is only identified up
to a possible sign switch, we solve this problem a posteriori, by
running our PMCMC sampler and identify the factor loading signs afterwards.

\section{Proposed PMCMC algorithm\label{sec:Proposed-PMCMC-algorithm}}

\subsection{Preliminaries}

If we let $\mathcal{F}_{t-1}$ denote the history of the $\left\{ y_{t}\right\} $
process up to time $t-1$, and $p\left(h_{1t},h_{2t}|\mathcal{F}_{t-1},\varTheta\right)$
the density of latent variables $h_{1t}$ and $h_{2t}$ conditioned
on $\left(\mathcal{F}_{t-1},\varTheta\right)$, then the likelihood
function of $\varTheta$ given the data $y=\left(y_{1},...,y_{T}\right)$
is
\begin{eqnarray}
p\left(y|\varTheta\right) & = & \prod_{t=1}^{T}\int p\left(y_{t}|h_{1t},h_{2t},\varTheta\right)p\left(h_{1t},h_{2t}|\mathcal{F}_{t-1},\varTheta\right)dh_{1t}dh_{2t}\nonumber \\
 & = & \prod_{t=1}^{T}\int N_{p}\left(y_{t}|RV_{t}^{1/2}T_{1}^{-1/2}\eta_{1t},\Omega_{t}\right)p\left(h_{1t},h_{2t}|\mathcal{F}_{t-1},\psi\right)dh_{1t}dh_{2t},
\end{eqnarray}
where $R=diag\left\{ \rho_{1},...,\rho_{p}\right\} $, $T_{1}=diag\left\{ \tau_{11}^{2},...,\tau_{1p}^{2}\right\} $,
$N_{p}\left(.|.,.\right)$ is the multivariate normal density function
with the mean $RV_{t}^{1/2}T_{1}^{-1/2}\eta_{1t}$ marginalised over
$f_{t}$, with the variance given by
\[
\Omega_{t}=BD_{t}B^{'}+V_{t}-R^{2}V_{t}.
\]
It is clear to see that neither $p\left(h_{1t},h_{2t}|\mathcal{F}_{t-1},\psi\right)$
nor the integral of $N_{p}\left(y_{t}|RV_{t}^{1/2}T_{1}^{-1/2}\eta_{1t},\Omega_{t}\right)$
over $\left(h_{1t},h_{2t}\right)$ are available in closed form. We
utilise PMCMC algorithm to develop a novel Bayesian estimation approach
for this model. Firstly, we discuss sampling the factor loading matrix
$B$ and the latent factors $\boldsymbol{f}=\left\{ \boldsymbol{f}_{j,.}\right\} $,
$j=1,...,k$, where $\boldsymbol{f}_{j,.}=\left(f_{j1},...,f_{jT}\right)^{'}$.
Conditional on knowing $\boldsymbol{h}_{1}$, $\boldsymbol{h}_{2}$,
and $\boldsymbol{f}$, the $B$ can be sampled conditionally on each
other from the multivariate normal distribution similar to a standard
factor model \citep{Lopes2004}. Sampling the factor loadings $B_{s,.}^{'}$
for $s=1,...,p$, conditionally on $\boldsymbol{f}$ from the conditional
posterior density $\widetilde{\pi}\left(B_{s,.}^{'}|\boldsymbol{f},\boldsymbol{y}_{s,.},\boldsymbol{h}_{1s,.}\right)$
can be done independently for each $s$, by performing a Gibbs-update
from
\[
B_{s,.}^{'}|\boldsymbol{f},\boldsymbol{y}_{s,.},\boldsymbol{h}_{1s,.}\sim N_{k_{s}}\left(a_{sT},b_{sT}\right),
\]
where $b_{sT}=\left(F_{s}^{'}V_{s}^{-1}F_{s}+B_{0}^{-1}I_{k_{s}}\right)^{-1}$
and $a_{sT}=b_{sT}F_{s}^{'}\left(V_{s}^{-1}\boldsymbol{y}_{s,.}-V_{s}^{-1}\frac{\rho_{s}}{\tau_{1s}}\exp\left(\boldsymbol{h}_{1s,.}/2\right)\eta_{1s,.}\right)$,
\[
F_{s}=\left[\begin{array}{ccc}
f_{11} & \cdots & f_{k_{i}1}\\
\vdots &  & \vdots\\
f_{1T} & \cdots & f_{k_{i}T}
\end{array}\right]
\]
and
\[
V_{s}=\left[\begin{array}{ccc}
\exp\left(h_{1s,1}\right)\left(1-\rho_{s}^{2}\right) & \cdots & 0\\
0 & \ddots & 0\\
0 & \cdots & \exp\left(h_{1s,T}\right)\left(1-\rho_{s}^{2}\right)
\end{array}\right].
\]
Sampling of $\left\{ f_{t}\right\} $: The sampling of the factors
are completed by sampling $\left\{ f_{t}\right\} $ from the distribution
$\left\{ f_{t}\right\} |y,\left\{ h_{1t}\right\} ,\left\{ h_{2t}\right\} ,B$.
After completing some algebra, we can show that $\left\{ f_{t}\right\} $
can be sampled from Gaussian with variance $b_{t}=\left(B^{'}\left(V_{t}-R^{2}V_{t}\right)^{-1}B+D_{t}^{-1}\right)^{-1}$
and mean  $a_{t}=b_{t}B^{'}\left[\left(V_{t}-R^{2}V_{t}\right)^{-1}y_{t}-\left(V_{t}-R^{2}V_{t}\right)^{-1}RV_{t}^{1/2}T_{1}^{-1/2}\eta_{1t}\right]$.

Sampling $\rho$, $\mu$, $\phi$, $\tau^{2}$, $\left\{ h_{1t}\right\} $,
and $\left\{ h_{2t}\right\} $: in the next step of the algorithm,
given $\left(y,B,f_{t}\right)$, and the conditional independence
of the errors, we exploit the fact that this models separates into
$p$ univariate SV models with leverage, and $k$ univariate SV models.
This shows that the latent idiosynchratic and factor volatilities
and SV specific parameters can be sampled series-by-series. This is
one of the reason that our approach is scalable in both $p$ and $k$.

\subsection{State Space Models and Sequential Monte Carlo Methods}

In this section, we first briefly describe state space model in general.
Let $\left\{ h_{t}\right\} _{t\in N}$ is a latent Markov process
with initial density $p_{\theta}\left(h_{1}\right)$ and state transition
density $p_{\theta}\left(h_{t}|h_{t-1}\right)$ for $t=1,...,T$.
The latent process $\left\{ h_{t}\right\} _{t\in N}$ is observed
only through $\left\{ y_{t}\right\} _{t\in N}$, whose value at time
$t$ depends on the value of hidden state at time $t$. This $p_{\theta}\left(y_{t}|h_{t}\right)$
is often called observation/measurement density. The joint probability
density function of $\left(h_{1:T},y_{1:T}\right)$ is
\[
p\left(h_{1:T},y_{1:T}|\theta\right)=p_{\theta}\left(h_{1}\right)p_{\theta}\left(y_{1}|h_{1}\right)\prod_{t=2}^{T}p_{\theta}\left(h_{t}|h_{t-1}\right)p_{\theta}\left(y_{t}|h_{t}\right).
\]
We also define the likelihood as $Z_{1:T}\left(\theta\right)=\prod_{t=1}^{T}Z_{t}\left(\theta\right)$,
where $Z_{1}\left(\theta\right)=p\left(y_{1}|\theta\right)$ and $Z_{t}\left(\theta\right)=p\left(y_{t}|y_{1:t-1},\theta\right)$.
The joint filtering density of $h_{1:t}$ can be written as
\[
\pi_{t}\left(h_{1:t}\right)=p\left(h_{1:t}|y_{1:t},\theta\right)=\frac{p\left(h_{1:t},y_{1:t}|\theta\right)}{Z_{1:t}\left(\theta\right)}
\]
and the joint posterior density of $\theta$ and $h_{1:T}$ is given
by
\[
p\left(\theta,h_{1:T}|y_{1:T}\right)=\frac{p\left(h_{1:T},y_{1:T}|\theta\right)p\left(\theta\right)}{\overline{Z}_{1:T}},
\]
where $\overline{Z}_{1:T}=\int_{\Theta}Z_{1:T}\left(\theta\right)p\left(\theta\right)d\theta$
is the marginal likelihood.

For a Bayesian analysis in a non-linear, non Gaussian state space
model, such as SV model with or without leverage, the ``ideal''
Gibbs sampler targeting the joint posterior density $p\left(\theta,h_{1:T}|y_{1:T}\right)$
consists of sampling alternately from the full conditional posteriors
$p_{\theta}\left(h_{1:T}|y_{1:T}\right)$ and $p\left(\theta|h_{1:T},y_{1:T}\right)$.
This is typically infeasible since exact sampling from $p_{\theta}\left(h_{1:T}|y_{1:T}\right)$
is impossible. The particle Gibbs approach of \citet{Andrieu:2010}
and the mixed sampler of \citet{Mendes2016} use a sequential Monte
Carlo (SMC) algorithm to obtain approximate samples from $p_{\theta}\left(h_{1:T}|y_{1:T}\right)$.

Sequential Monte Carlo (SMC) algorithms consist of recursively producing
a weighted particles $\left\{ h_{1:t}^{i},w_{t}^{i}\right\} _{i=1}^{N}$
such that the intermediate target density $p\left(h_{1:t}|y_{1:t},\theta\right)$
can be approximated by
\[
\pi_{t}\left(dh_{1:t}\right)=\sum_{i=1}^{N}W_{t}^{i}\delta_{h_{1:t}^{i}}\left(dh_{1:t}\right),\;\;W_{t}^{i}=\frac{w_{t}^{i}}{\sum_{l=1}^{N}w_{t}^{l}},
\]
where $\delta$ denote the Dirac delta mass located at $h$. Suppose
that at the end of period $\left(t-1\right)$, we have a set of particle
$\left\{ h_{1:t-1}^{i},W_{t-1}^{i}\right\} _{i=1}^{N}$. Once we have
a new observation $y_{t}$, we propagate the particles $h_{1:t-1}^{i}$
to $h_{1:t}^{i}=\left(h_{t}^{i},h_{1:t-1}^{i}\right)$ using the importance
sampling (IS) density $m\left(h_{t}^{i}|h_{1:t-1}^{i}\right)$ and
updating the corresponding importance sampling weights according to
\begin{equation}
w_{t}^{i}=W_{t-1}^{i}\frac{p\left(y_{t}|h_{t}^{i}\right)p\left(h_{t}^{i}|h_{t-1}^{i}\right)}{m\left(h_{t}^{i}|h_{1:t-1}^{i}\right)},\label{eq:ISweight}
\end{equation}
with the corresponding normalised weights calculated as $W_{t}^{i}=w_{t}^{i}/\sum_{i=1}^{N}w_{t}^{i}$.
The variance of IS weights $w_{t}^{i}$ in \eqref{eq:ISweight} increases
exponentially with the time period $t$ and hence reducing the effective
sample size in the particle filter. This is known as 'weight degeneracy'
problem. To avoid this problem, SMC algorithm needs to include a resampling
step before propagating the particles $h_{1:t-1}^{i}$ to $h_{1:t}^{i}=\left(h_{t}^{i},h_{1:t-1}^{i}\right)$.
The $N$ 'ancestor particles' from $\left\{ h_{1:t-1}^{i}\right\} _{i=1}^{N}$
is sampled according to their normalised IS weights $\left\{ W_{t-1}^{i}\right\} $
and then set the IS weights $W_{t-1}^{i}$ all equal to $1/N$. Popular
resampling schemes include multinomial, residual, stratified, and
systematic resampling.

Another issue in implementing SMC efficiently is the choice of the
IS densities $q_{t}\left(h_{t}|h_{1:t}^{i}\right)$. In general, this
requires to select $m_{t}\left(h_{t}|h_{1:t}^{i}\right)$ as a close
approximation to the period-$t$ conditional density $\pi_{t}\left(h_{t}|h_{1:t-1}^{i}\right)$.
The most popular selection for importance sampling densities are the
transition densities $p_{\theta}\left(h_{t}^{i}|h_{t-1}^{i}\right)$
used by Bootstrap Particle Filter \citep{Gordon:1993}. In the case
of the measurement density $p_{\theta}\left(y_{t}|h_{t}\right)$ is
quite flat in $h_{t}$, this selection typically sufficient. There
are more advances particle filter algorithms developed in the literature,
such as Particle efficient importance sampling of \citet{Scharth:2016}
and Auxiliary Particle filter of \citet{Pitt:1999} that are more
efficient than standard bootstrap particle filter.

The SV model with leverage can be expressed in the form of state space
model consisting of the measurement density
\begin{multline*}
p_{\theta}\left(y_{st}|h_{1st},h_{1st-1}\right)\sim N\left(B_{s}f_{t}+\frac{\rho_{1s}}{\tau_{1s}}\exp\left(h_{1st}/2\right)\left(h_{1st}-\mu_{1s}-\phi_{1s}\left(h_{1st-1}-\mu_{1s}\right)\right)\right.\\
\left.,\left(1-\rho_{1s}^{2}\right)\exp\left(h_{1st}\right)\right)
\end{multline*}
and this following state transition density

\[
p_{\theta}\left(h_{1st}|h_{1st-1}\right)\sim N\left(\mu_{1s}+\phi_{1s}\left(h_{1st-1}-\mu_{1s}\right),\tau_{1s}^{2}\right),for\;s=1,...,p,
\]
where, $p_{\theta}\left(h_{1i1}\right)\sim N\left(\mu_{1i},\frac{\tau_{1i}^{2}}{1-\phi_{1i}^{2}}\right)$.
For each $j=1,...,k$, the SV model can be expressed in the form of
state space model consisting of the measurement density

\[
p\left(f_{jt}|h_{2jt}\right)\sim N\left(0,\exp\left(h_{2jt}\right)\right)
\]
and this following state transition density

\[
p\left(h_{2jt}|h_{2jt-1}\right)\sim N\left(\phi_{2j}h_{2jt-1},\tau_{2j}^{2}\right),
\]
where $p_{\theta}\left(h_{2j1}\right)\sim N\left(\mu_{2j},\frac{\tau_{2j}^{2}}{1-\phi_{2j}^{2}}\right)$.

\subsection{Target Distribution of Particle Markov chain Monte Carlo (PMCMC)\label{sub:Target-Distribution-of}}

In this section, we first define the appropriate target density for
factor MSV that include all the random variables which are produces
by SMC to generate $h_{1s,1:T}$ for $s=1,...p$ and $h_{2j,1:T}$
for $j=1,...,r$. We first approximate the joint filtering densities
$\left\{ p\left(h_{1st}|y_{s,1:t},\varTheta\right):t=1,...,T\right\} $
for $s=1,...,p$ and $\left\{ p\left(h_{2jt}|f_{j,1:t},\varTheta\right):t=1,...,T\right\} $
sequentially, using particles, i.e. weighted samples $\left(h_{1st}^{1:N},\overline{w}_{1st}^{1:N}\right)$
and $\left(h_{2jt}^{1:N},\overline{w}_{2jt}^{1:N}\right)$, drawn
from some important densities $m_{1st}^{\varTheta}$ and $m_{2jt}^{\varTheta}$
for $t=1,...,T$, respectively. A valid resampling scheme $M_{1}\left(a_{1st-1}^{1:N}\right)$,
where each $a_{1st-1}^{i}=k$ indexed a particle in $\left(h_{1st}^{1:N},\overline{w}_{1st}^{1:N}\right)$
and is chosen with probability $\overline{w}_{1st}^{k}$, $M_{2}\left(a_{1jt-1}^{1:N}\right)$
is defined similarly. The Sequential Monte Carlo algorithm used in
this paper is the same as in \citet{Andrieu:2010} and it is given
in the Appendix \ref{sec:Generic-Sequential-Monte}. We denote the
vector of particles by
\[
U_{1s,1:T}\coloneqq\left(h_{1s,1}^{1:N},...,h_{1s,T}^{1:N},C_{1s,1}^{1:N},...,C_{1s,T-1}^{1:N}\right)
\]
and
\[
U_{2j,1:T}\coloneqq\left(h_{2j,1}^{1:N},...,h_{2j,T}^{1:N},C_{2j,1}^{1:N},...,C_{2j,T-1}^{1:N}\right).
\]
This SMC algorithm also provides an estimate of the likelihood $Z_{1s}\left(U_{1s,1:T},\varTheta\right)=\prod_{t=1}^{T}\left(\frac{1}{N}\sum_{i=1}^{N}w_{1s,t}^{i}\right)$,
for $s=1,..,p$ and $Z_{2j}\left(U_{2j,1:T},\varTheta\right)=\prod_{t=1}^{T}\left(\frac{1}{N}\sum_{i=1}^{N}w_{2j,t}^{i}\right)$
for $j=1,..,k$. The joint distribution of the particles given the
parameters are
\begin{multline*}
\psi_{1s}\left(U_{1s,1:T}^{1:N}|\varTheta\right)\coloneqq\\
\prod_{i=1}^{N}m_{1s1}^{\varTheta}\left(h_{1s1}^{i}\right)\prod_{t=2}^{T}\left\{ M\left(a_{1st-1}^{1:N}|\overline{w}_{1st-1}^{1:N}\right)\prod_{i=1}^{N}m_{1st}^{\varTheta}\left(x_{1st}^{i}|x_{1st-1}^{a_{1st-1}^{i}}\right)\right\}
\end{multline*}
for $s=1,...,p$ and
\begin{multline*}
\psi_{2j}\left(U_{2j,1:T}^{1:N}|\varTheta\right)\coloneqq\\
\prod_{i=1}^{N}m_{2j1}^{\varTheta}\left(h_{2j1}^{i}\right)\prod_{t=2}^{T}\left\{ M\left(a_{2jt-1}^{1:N}|\overline{w}_{2jt-1}^{1:N}\right)\prod_{i=1}^{N}m_{2jt}^{\varTheta}\left(x_{2jt}^{i}|x_{2jt-1}^{a_{2jt-1}^{i}}\right)\right\}
\end{multline*}
for $j=1,...,k$. We then construct a target distribution on an augmented
space that includes the particles $U_{1s,1:T}^{1:N}$, for $s=1,...,p$
and $U_{2j,1:T}^{1:N}$ for $j=1,...,k$.

In this paper, we use simple ancestral tracing method of \citet{Kitagawa:1996}
to sample one particle from the final particle filter. The method
is equivalent to sampling index $J_{1s}=j_{1s}$, for $s=1,...,P$,
with probability $\overline{w}_{T}^{j_{1s}}$, tracing back its ancestral
lineage $C_{1s,1:T-1}^{J_{1s}}$ and choosing the particle $h_{1s,1:T}^{j_{1s}}=\left(h_{1s,1}^{C_{1s1}^{j_{1s}}},...,h_{1s,T}^{C_{1sT}^{j_{1s}}}\right)$,
and $h_{2j,1:T}^{j_{2j}}$ can be obtained similarly. Further, let
us denote
\[
u_{1s,1:T}^{\left(-j_{1s}\right)}=\left\{ h_{1s1}^{\left(-C_{1s1}^{j_{1s}}\right)},...,h_{1s,T-1}^{\left(-C_{1sT-1}^{j_{1s}}\right)},h_{1s,T}^{\left(-j_{1s}\right)},a_{1s,1}^{\left(-C_{1s1}^{j_{1s}}\right)},...,a_{1sT-1}^{\left(-C_{1sT-1}^{j_{1s}}\right)}\right\} .
\]
Then, the target distribution is given by
\begin{multline}
\tilde{\pi}^{N}\left(h_{1,1:T},h_{2,1:T},C_{1,1:T-1},C_{2,1:T-1},J_{1},J_{2},U_{1,1:T}^{\left(-J_{1}\right)},U_{2,1:T}^{\left(-J_{2}\right)},\varTheta,f\right)\coloneqq\\
\frac{p\left(h_{1,1:T},h_{2,1:T},\varTheta,f|y_{1:T}\right)}{N^{T}}\prod_{s=1}^{P}\frac{\psi_{1s}\left(U_{1s,1:T}|\varTheta\right)}{m_{1s1}^{\varTheta}\left(h_{1s1}^{C_{1s1}}\right)\prod_{t=2}^{T}\tilde{w}_{1s,t-1}^{a_{1st-1}^{C_{1st}}}m_{1st}^{\theta}\left(h_{1st}^{C_{1st}}|h_{1st-1}^{a_{1st-1}^{C_{1st}}}\right)}\\
\prod_{j=1}^{k}\frac{\psi_{2j}\left(U_{2j,1:T}|\varTheta\right)}{m_{2j1}^{\varTheta}\left(h_{2j1}^{C_{2j1}}\right)\prod_{t=2}^{T}\tilde{w}_{2j,t-1}^{a_{2jt-1}^{C_{2jt}}}m_{2jt}^{\theta}\left(h_{2jt}^{C_{2jt}}|h_{2jt-1}^{a_{2jt-1}^{C_{2jt}}}\right)}.\label{eq:target distribution}
\end{multline}
Assumption 1 of \citet{Andrieu:2010} ensures that the SMC approximation
$\psi_{1s}\left(U_{1s,1:T}|\varTheta\right)$ and $\psi_{2j}\left(U_{2j,1:T}|\varTheta\right)$
can be used as a Metropolis-Hasting proposal density for generating
from $\tilde{\pi}^{N}\left(U_{1s,1:T}|\varTheta\right)$ for $s=1,...,p$
and $\tilde{\pi}^{N}\left(U_{2j,1:T}|\varTheta\right)$ for $j=1,...,k$.
Equation \eqref{eq:target distribution} has the following marginal
distribution

\begin{alignat*}{1}
 & \tilde{\pi}^{N}\left(h_{1,1:T},h_{2,1:T},C_{1,1:T-1},C_{2,1:T-1},J_{1},J_{2},\varTheta,f\right)\\
 & \coloneqq\int\int\tilde{\pi}^{N}\left(h_{1,1:T},h_{2,1:T},C_{1,1:T-1},C_{2,1:T-1},J_{1},J_{2},U_{1,1:T}^{\left(-J_{1}\right)},U_{2,1:T}^{\left(-J_{2}\right)},\varTheta,f\right)dU_{1,1:T}^{\left(-J_{1}\right)}dU_{2,1:T}^{\left(-J_{2}\right)}\\
 & \coloneqq\frac{p\left(h_{1,1:T}^{J_{1}},h_{2,1:T}^{J_{2}},\varTheta,f|y_{1:T}\right)}{N^{T}}.
\end{alignat*}
This is defined to be the target density of interest $p\left(h_{1,1:T}^{J_{1}},h_{2,1:T}^{J_{2}},\varTheta,f|y_{1:T}\right)$
up to the factor $1/N^{T}$ representing a discrete uniform density
over the index variables in $\left(C_{1,1:T-1},C_{2,1:T-1}\right)$
and hence
\[
\tilde{\pi}^{N}\left(h_{1,1:T},h_{2,1:T},\varTheta,f\right)\coloneqq p\left(h_{1,1:T},h_{2,1:T},\varTheta,f|y_{1:T}\right).
\]
It follows that Pseudo Marginal, particle Gibbs, and mixed samplers
leaves the target density $p\left(h_{1,1:T},h_{2,1:T},\varTheta,f|y_{1:T}\right)$
invariant and delivers under weak reqularity conditions a sequence
of draws $\left\{ h_{1,1:T}^{\left(l\right)},h_{2,1:T}^{\left(l\right)},\varTheta^{\left(l\right)},f^{\left(l\right)}\right\} $
whose marginal distributions converge for any $N>1$ to $p\left(h_{1,1:T},h_{2,1:T},\varTheta,f|y_{1:T}\right)$
as $l\rightarrow\infty$ \citep{Andrieu:2010}.

\subsection{PMCMC (Particle Gibbs and Mixed) Sampling Schemes}

This section describes a version of Particle Gibbs (PG) of \citet{Andrieu:2010}
and mixed sampling schemes of \citet{Mendes2016} for Factor MSV using
the target distributions given in Section \ref{sub:Target-Distribution-of}.

\subsubsection{Particle Gibbs (PG) and Particle Metropolis within Gibbs (PMwG) \label{sub:Particle-Gibbs algorithm}}

The Particle Gibbs is a standard Gibbs sampler for the augmented target
distribution in equation \eqref{eq:target distribution}. The Gibbs
sampler for this augmented density requires a different type of SMC
algorithm, referred to as conditional SMC, where one of the particles
is specified a priori. This reference particle denoted by $\left(h_{1,1:T}^{J_{1}},h_{2,1:T}^{J_{2}}\right)$
is then retained throughtout the entire SMC sampling process \citep{Andrieu:2010}.
To accomplish this, we also need special resampling schemes. We use
conditional systematic resampling in \citet{Chopin:2013}. The conditional
SMC algorithm is given in Appendix \ref{sec:Conditional-Sequential-Monte}.

The PG or PMwG sampling schemes for Factor MSV with leverage in Section
\ref{sec:Factor-SV-Model} can be proceed as follows:
\begin{enumerate}
\item Loop over $\theta_{1s}$: $\left(\mu_{1s},\phi_{1s},\rho_{1s},\tau_{1s}^{2}\right)$,
for each $s=1,...,p$

\begin{enumerate}
\item Sample from conditional distribution $p\left(\theta_{1s}|h_{1,1:T}^{J_{1}},C_{1,1:T}^{J_{1}},J_{1},\theta_{-1s},\theta_{1s},f,B\right)$
if it is available.
\item Or, sample $\theta_{1s}^{*}\sim q_{1s}\left(.|h_{1,1:T}^{J_{1}},C_{1,1:T}^{J_{1}},J_{1},\theta_{-1s},\theta_{1s},f,B\right)$
\item Accept with probability
\begin{multline*}
\alpha\left(\theta_{1s}^{*},\theta_{1s}|h_{1s,1:T}^{J_{1s}},C_{1s,1:T}^{J_{1s}},J_{1s},\theta_{-1s},f,B\right)=1\wedge\frac{\pi\left(\theta_{1s}^{*}|h_{1s1:T}^{J_{1s}},C_{1s,1:T-1}^{J_{1s}},J_{1s},\theta_{-1s},f,B\right)}{\pi\left(\theta_{1s}|h_{1s,1:T}^{J_{1s}},C_{1s,1:T-1}^{J_{1s}},J_{1s},\theta_{-1s},f,B\right)}\times\\
\frac{q_{s}\left(\theta_{1s}|h_{1s,1:T}^{J_{1s}},C_{1s,1:T-1}^{J_{1s}},J_{1s},\theta_{-1s},\theta_{1s}^{*},f,B\right)}{q_{s}\left(\theta_{1s}^{*}|h_{1s,1:T}^{J_{1s}},C_{1s,1:T-1}^{J_{1s}},J_{1s},\theta_{-1s},\theta_{1s},f,B\right)}.
\end{multline*}

\end{enumerate}
\item Loop over $\theta_{2j}:\left(\phi_{2j},\tau_{2j}^{2}\right)$ , for
each $j=1,...,k$

\begin{enumerate}
\item Sample from conditional distribution $p\left(\theta_{2j}|h_{2,1:T}^{J_{2}},C_{2,1:T}^{J_{2}},J_{2},\theta_{-2j},\theta_{2j},f,B\right)$
if it is available.
\item Sample $\theta_{2j}^{*}\sim q_{2j}\left(.|h_{2,1:T}^{J_{2j}},C_{2,1:T}^{J_{2j}},J_{2j},\theta_{-2j},\theta_{2j},f,B\right)$
\item Accept with probability
\begin{multline*}
\alpha\left(\theta_{2j}^{*},\theta_{2j}|h_{2j,1:T}^{J_{2j}},C_{2j,1:T}^{J_{2j}},J_{2j},\theta_{-2j},f,B\right)=1\wedge\frac{\pi\left(\theta_{2j}^{*}|h_{2j,1:T}^{J_{2j}},C_{2j,1:T-1}^{J_{2j}},J_{2j},\theta_{-2j},f,B\right)}{\pi\left(\theta_{2j}|h_{2j,1:T}^{J_{2j}},C_{2j,1:T-1}^{J_{2j}},J_{2j},\theta_{-2j},f,B\right)}\times\\
\frac{q_{j}\left(\theta_{2j}|h_{2j,1:T}^{J_{2j}},C_{2j,1:T-1}^{J_{2j}},J_{2j},\theta_{-2j},\theta_{2j}^{*},f,B\right)}{q_{j}\left(\theta_{2j}^{*}|h_{2j,1:T}^{J_{2j}},C_{2j,1:T-1}^{J_{2j}},J_{2j},\theta_{-2j},\theta_{2j},f,B\right)}
\end{multline*}

\end{enumerate}
\item Generate $B$ from $\pi\left(B|h_{1,1:T}^{J_{1}},C_{1,1:T-1}^{J_{1}},J_{1},h_{2,1:T}^{J_{2}},C_{2,1:T-1}^{J_{2}},J_{2},\varTheta_{-\beta},\boldsymbol{f}\right)$.
\item Generate $f_{t}$ for $t=1,..,T$ from $\pi\left(f_{t}|\left\{ h_{1t}\right\} ,\left\{ h_{2t}\right\} ,B,\varTheta\right)$.
\item For $s=1,...,p$, sample $U_{1s,1:T}^{\left(-J_{1s}\right)}\sim\pi\left(U_{1s,1:T}^{-J_{1s}}|h_{1s,1:T}^{J_{1s}},C_{1s,1:T-1}^{J_{1s}},J_{1s},\varTheta\right)$,
this is the conditional sequential Monte Carlo step, in which a particle
$h_{1s,1:T}^{J_{1s}}$ and the associated sequence of ancestral indices
$C_{1s,1:T-1}^{J_{1s}}$ are kept unchanged. The conditional sequential
Monte Carlo is a procedure that resamples all the particles and indices
except for $U_{1s,1:T}^{J_{1s}}$.
\item For $s=1,...,p$, sample $J_{1s}\sim\pi\left(J_{1s}|U_{1s,1:T},\varTheta\right)$
\item For $j=1,...,k$, sample $U_{2j,1:T}^{\left(-J_{2j}\right)}\sim\pi\left(U_{2j,1:T}^{-J_{2j}}|h_{2j,1:T}^{J_{2j}},C_{2j,1:T-1}^{J_{2j}},J_{2j},\varTheta\right)$,
this is conditional sequential Monte Carlo step and is given in Appendix
\ref{sec:Conditional-Sequential-Monte}.
\item For $j=1,...,k$, sample $J_{2j}\sim\pi\left(J_{2j}|U_{2j,1:T},\varTheta\right)$.
\end{enumerate}
As is known in the literature that this PG or PMwG implemented using
bootstap particle filter with resampling steps at every period of
$t$, have a very poor mixing, especially when the time period $T$
is large. This is due to path degeneracy problem \citep{Lindsten:2014}.
The consequence of this path degeneracy problem is that at iteration
step $l$ the new path trajectory $h_{1:T}^{\left(l\right)}$ tend
to coalesce with the previous one $h_{1:T}^{\left(l-1\right)}$ which
is retained as the reference particle trajectory in conditional sequential
Monte Carlo sampling. The resulting particle degenerate toward this
reference trajectory, and leads to poor mixing Markov chain.

In order to address the mixing problem of the PG caused by path degeneracy,
we add additional Ancestor Sampling steps to conditional SMC (PGAS),
which assign at each time period $t$ a new artificial $h_{1:t-1}$
history to the reference path $h_{t:T}^{J}$. The PGAS augments each
period-$t$ conditional SMC resampling step by randomly selecting
from the set $\left\{ h_{1:t-1}^{i}\right\} _{i=1}^{N}$ (including
the reference trajectory) one ancestor particle which is used as a
new history to the partially reference trajectory $h_{t:T}^{J}$.
In \citet{Lindsten:2014}, the PGAS is implemented using bootstrap
particle filter, the ancestor sampling weights are given by
\[
\bar{w}_{t-1|T}\propto w_{t-1}^{i}p_{\theta}\left(h_{t}^{J_{t}}|h_{t-1}^{i}\right).
\]
\citet{Lindsten:2014} shows that the invariance property of PG is
not violated by this additional ancestor sampling step. Because this
ancestor sampling step assign a new ancestor to $h_{t:T}^{J}$ in
each period, then it will produce new trajectory $h_{t:T}^{'}$ that
tends to be different from the reference trajectory $h_{t:T}^{J}$.

\subsubsection{Mixed Sampling Schemes}

\citet{Mendes2016} proposed a mixed PMCMC sampler which combine the
PG and pseudo marginal method. This mixed sampler is highly efficient
when there is a set of parameters that is not highly correlated with
the latent states which can be generated using PG, and another set
of parameters that is highly correlated with the latent states and
is generated using the PMMH sampler. After some experimentation with
univariate SV, we found that $PMMH\left(\tau^{2}\right)+PG\left(\mu,\phi\right)$
is the most efficient sampler. by using their notation, for Factor
MSV model, we follow
\[
PMMH\left(\tau_{11}^{2},...,\tau_{1p}^{2},\tau_{21}^{2},...,\tau_{2k}^{2}\right)+PG\left(\mu_{1s},\phi_{1s},\rho_{1s},s=1,...,p;\mu_{1j},\phi_{1j},j=1,...,k;B,f\right).
\]
The sampling scheme is given by:
\begin{enumerate}
\item Pseudo Marginal step for $\tau_{1s}^{2}$, For $s=1,...,P$

\begin{enumerate}
\item Sample $\tau_{1s}^{2*}\sim q_{1s,\tau^{2}}\left(.|U_{1,1:T},U_{2,1:T},J_{1},\theta_{-\tau_{1s}^{2}},f,B\right)$
\item Sample $U_{1s,1:T}^{*}\sim\psi\left(.|U_{1_{-s},1:T},U_{2,1:T},\theta_{-\tau_{1s}^{2}},\tau_{1s}^{2*},f,B\right)$
\item Sample $J_{1s}^{*}$ from $\pi\left(.|U_{1s,1:T}^{*},U_{1_{-s},1:T},U_{2,1:T},\theta_{-\tau_{1s}^{2}},\tau_{1s}^{2*},f,B\right)$
\item Accept with probability:
\begin{multline}
\alpha\left(U_{1s,1:T},J_{1s},\tau_{1s}^{2};U_{1s,1:T}^{*},J_{1s}^{*},\tau_{1s}^{2*}|f,\theta_{-\tau_{1s}^{2}},B\right)=1\wedge\frac{Z\left(\theta_{-\tau_{1s}^{2}},\tau_{1s}^{2*},U_{1s,1:T}^{*}\right)}{Z\left(\theta_{-\tau_{1s}^{2}},\tau_{1s}^{2},U_{1s,1:T}\right)}\times\\
\frac{q_{1s,\tau^{2}}\left(\tau_{1s}^{2}|U_{1s,1:T}^{*},J_{1s}^{*},\theta_{-\tau_{1s}^{2}},f,\tau_{1s}^{2*},B\right)}{q_{1s,\tau^{2}}\left(\tau_{1s}^{2*}|U_{1s,1:T},J_{1s},\theta_{-\tau_{1s}^{2}},f,\tau_{1s}^{2},B\right)}\times\frac{p\left(\tau_{1s}^{2*}|\theta_{-\tau_{1s}^{2}}\right)}{p\left(\tau_{1s}^{2}|\theta_{-\tau_{1s}^{2}}\right)}
\end{multline}

\end{enumerate}
\item Pseudo Marginal step for $\tau_{2j}^{2}$, for $j=1,...,k$

\begin{enumerate}
\item Sample $\tau_{2j}^{2*}\sim q_{2j,\tau^{2}}\left(.|U_{1,1:T},U_{2,1:T},J_{2},\theta_{-\tau_{2j}^{2}},f,B\right)$
\item Sample $U_{2j,1:T}^{*}\sim\psi\left(.|U_{1,1:T},U_{2_{-j},1:T},\theta_{-\tau_{2j}^{2}},\tau_{2j}^{2*},f,B\right)$
\item Sample $J_{2j}^{*}$ from $\pi\left(.|U_{1,1:T},U_{2_{-j},1:T},U_{2j,1:T}^{*},\theta_{-\tau_{2j}^{2}},\tau_{2j}^{2*},f,B\right)$
\item Accept with probability:
\begin{multline}
\alpha\left(U_{2j,1:T},J_{2j},\tau_{2j}^{2};U_{2j,1:T}^{*},J_{2j}^{*},\tau_{2j}^{2*}|f,\theta_{-\tau_{2j}^{2}},B\right)=1\wedge\frac{Z\left(\theta_{-\tau_{2j}^{2}},\tau_{2j}^{2*},U_{2j,1:T}^{*}\right)}{Z\left(\theta_{-\tau_{2j}^{2}},\tau_{2j}^{2},U_{2j,1:T}\right)}\times\\
\frac{q_{2j,\tau^{2}}\left(\tau_{2j}^{2}|U_{2j,1:T}^{*},J_{2j}^{*},\theta_{-\tau_{2j}^{2}},f,\tau_{2j}^{2*},B\right)}{q_{2j,\tau^{2}}\left(\tau_{2j}^{2*}|U_{2j,1:T},J_{2j},\theta_{-\tau_{2j}^{2}},f,\tau_{2j}^{2},B\right)}\times\frac{p\left(\tau_{2j}^{2*}|\theta_{-\tau_{2j}^{2}}\right)}{p\left(\tau_{2j}^{2}|\theta_{-\tau_{2j}^{2}}\right)}
\end{multline}

\end{enumerate}
\item Followed by step $1$ to $8$ of PG algorithm, except that the $\tau_{1s}^{2}$,
$s=1,...,p$ and $\tau_{2j}^{2}$, $j=1,...,k$ are not generated
in step 1 and 2 of PG algorithm.
\end{enumerate}
Note that part 3 is the same as PG or Particle Metropolis within Gibbs
algorithm described in Section \ref{sub:Particle-Gibbs algorithm}.
Part 1 and 2 also generates the variable $J_{1}$ and $J_{2}$ which
select the trajectory for each series and factors. This is necessary
since $J_{1}$ and $J_{2}$ are used in the PG/PMwG step.

\subsection{Prior Distributions}

To perform Bayesian inference, the prior distributions for the parameters
need to be specified. Independently, for each $s=1,...,p$, priors
for the idiosynchratic SV parameters $p\left(\mu_{1s},\phi_{1s},\tau_{1s}^{2}\right)=p\left(\mu_{1s}\right)p\left(\phi_{1s}\right)p\left(\tau_{1s}^{2}\right)$,
where the prior for $p\left(\mu_{1s}\right)\propto1$, the prior for
the persistence parameter $\phi_{s}\in\left(-1,1\right)$ follows
$U\left(-1,1\right)$, and the prior for $\tau$ follow half-cauchy
distribution such that the prior for $\tau^{2}$ is given by
\[
p\left(\tau^{2}\right)=\frac{I\left(\tau>0\right)}{\pi\left(1+\tau^{2}\right)\sqrt{\tau^{2}}}.
\]
The prior for $\rho_{s}$ for $s=1,...,p$ is $U\left(-1,1\right)$.
Same prior is used for factor SV parameters $\left(\mu_{2j},\phi_{2j},\tau_{2j}^{2}\right)$
for $j=1,..,k$. The initial state $h_{1s,1}$ and $h_{2j,1}$ are
distributed according to the stationary distribution of the AR(1)
process, i. e. $h_{1s,1}|\mu_{1s},\phi_{1s},\tau_{1s}^{2}\sim N\left(\mu_{1s},\tau_{1s}^{2}/\left(1-\phi_{1s}^{2}\right)\right)$
and $h_{2j,1}|\mu_{2j},\phi_{2j},\tau_{2j}^{2}\sim N\left(\mu_{2j},\tau_{2j}^{2}/\left(1-\phi_{2j}^{2}\right)\right)$.
For every unrestricted element of the factor loadings matrix $B$,
we choose independent Gaussian distributions, i. e. $p\left(B_{sj}\right)\sim N\left(0,1\right)$.

\subsection{Sampling Factor Loading $B$ using Interweaving Method}

It is well-known that sampling factor loading $B$ conditioned on
$\left\{ f_{t}\right\} $ and then sampling $\left\{ f_{t}\right\} $
conditioned on $B$ is very inefficient and leads to extremely slow
convergence and poor mixing. To overcome this problem, \citet{Chib2006}
sample the factor loading matrix $B$ from $p\left(B|h_{1},h_{2},\varTheta_{-B}\right)$
without conditioning on the factor $\boldsymbol{f}$. However, without
conditioning on the factor $\boldsymbol{f}$, the full conditional
distribution is not available in closed form and to sample from it
requires Metropolis-Hastings update with high dimensional and complex
proposal that is based on numerically maximising the conditional posterior
and approximate the hessian of log-posterior at MCMC iteration. In
this paper we employ simpler approach based on an ancillarity-sufficiency
interweaving strategy (ASIS), in particular deep interweaving strategy,
introduced by \citet{Kastner:2017}. We briefly describe the deep
interweaving strategy.

The parameterisation underlying deep interweaving is given by
\begin{equation}
y_{t}=B^{*}f_{t}^{*}+u_{t},\;\;f_{t}^{*}|h_{2j,.}^{*}\sim N_{k}\left(0,diag\left(e^{h_{21,t}^{*}},...,e^{h_{2k,t}^{*}}\right)\right),\label{eq:deep interweaving factor model}
\end{equation}
with a lower triangular factor loading matrix $B^{*}$ where $B_{11}^{*}=1,...,B_{kk}^{*}=1$.
The factor model in equation\eqref{eq:standard factor model} can
be reparameterised into factor model in equation \eqref{eq:deep interweaving factor model}
using a simple linear transformation
\[
f_{t}^{*}=Df_{t},B^{*}=BD^{-1}
\]
for $t=1,..,T$. The $k$ latent factor volatilities $h_{2j,t}^{*}$
follow alternative univariate SV models with the level $\mu_{2j}=\log B_{jj}^{2}$
rather than zero as in factor SV model in Section \ref{sec:Factor-SV-Model}.
The transformation of the factor volatilities is given by
\[
h_{2j,t}^{*}=h_{2j,t}+\log B_{jj}^{2},t=0,...,T,j=1,...,k
\]
In between step 3 and 4 of the PG and step 5 and 6 of the mixed sampler,
we add this following deep interweaving algorithm and perform these
steps independently for each $j=1,..,k$
\begin{itemize}
\item Determine the vector $B_{.,j}^{*}$, where $B_{sj}^{*}=B_{sj}^{old}/B_{jj}^{old}$
in the $j$th column of the transformed factor loading matrix $B^{*}$.
\item Define $h_{2j,.}^{*}=h_{2j,.}^{old}+2\log|B_{jj}^{old}|$ and sample
$B_{jj}^{new}$ from $p\left(B_{jj}|B_{.,j}^{*},h_{2j,.}^{*},\phi_{2j},\tau_{2j}^{2}\right)$,
see \ref{sec:deep interweaving} and \citet{Kastner:2017} for details.
\item Update $B_{.,j}=\frac{B_{jj}^{new}}{B_{jj}^{old}}B_{.,j}^{old}$,
$f_{j,.}=\frac{B_{jj}^{old}}{B_{jj}^{new}}f_{j,.}^{old}$, and $h_{2j,.}=h_{2j,.}^{old}+2\log|\frac{B_{jj}^{old}}{B_{jj}^{new}}|$.
\end{itemize}

\subsection{Normal-Gamma prior distribution for factor loading matrix $B$}

The standard prior for each element of the factor loading matrix $B$
is a independent zero-mean normal distribution, $N\left(0,\sigma^{2}=1\right)$
for each $s=1,...,p$ and $j=1,...,k$. Following \citet{Griffin:2010},
we model the variance each variance $\sigma_{sj}^{2}$ as a random
variable and placing hyperprior on $\sigma_{sj}^{2}$ as follows

\[
B_{sj}|\sigma_{sj}^{2}\sim N\left(0,\sigma_{sj}^{2}\right),\;\;\sigma_{sj}^{2}|\lambda_{s}^{2}\sim G\left(a_{s},\lambda_{s}^{2}/2\right).
\]
We let $\lambda_{s}^{2}\sim G\left(c_{s},d_{s}\right)$, where $c_{s}$
and $d_{s}$ are fixed hyperparameters. The choice of $a$ and $\lambda$
plays an important role in the estimation. As the shape parameter
$a_{s}$ decreases these include distributions that place a lot of
mass close to zero, and at the same time heavy tails. This implies
that choosing small $a_{s}$ imposes strong shrinkage towards zero,
and choosing large $a_{s}$ imposes a little shrinkage towards zero.
The Bayesian Lasso prior of \citet{Park:2008} is a special case when
$a_{s}=1$.

After completing some algebra, for $s=1,...,p$, sample the full conditional
distribution of $\lambda_{s}^{2}|\sigma_{s.}^{2}$ from
\[
\lambda_{s}^{2}|\sigma_{s.}^{2}\sim G\left(c_{s}+a_{s}k_{s},d_{s}+\frac{1}{2}\sum_{s=1}^{k_{s}}\sigma_{sj}^{2}\right)
\]
where $k_{s}=\min\left(s,k\right)$, and then sample the full conditional
distribution of $\sigma_{sj}^{2}|\lambda_{s}^{2},B_{sj}$ from
\[
\sigma_{sj}^{2}|\lambda_{s}^{2},B_{sj}\sim GIG\left(a_{s}-\frac{1}{2},\lambda_{s}^{2},B_{sj}^{2}\right),
\]
where the generalised inverse Gaussian $GIG\left(m,k,l\right)$ distribution
has a density proportional to
\[
x^{m-1}\exp\left\{ -\frac{1}{2}\left(kx+l/x\right)\right\} .
\]
Let $\Psi_{s}=diag\left(\sigma_{i1}^{-2},\sigma_{i2}^{-2},...,\sigma_{ik_{s}}^{-2}\right)$,
we can draw
\[
B_{s,.}^{'}|\boldsymbol{f},\boldsymbol{y}_{s,.},\boldsymbol{h}_{1s,.}\sim N_{k_{s}}\left(a_{sT},b_{sT}\right),
\]
where $b_{sT}=\left(F_{s}^{'}V_{s}^{-1}F_{s}+\Psi_{s}\right)^{-1}$
and $a_{sT}=b_{sT}F_{s}^{'}\left(V_{s}^{-1}\boldsymbol{y}_{s,.}-V_{s}^{-1}\frac{\rho_{s}}{\tau_{1s}}\exp\left(\boldsymbol{h}_{1s,.}/2\right)\eta_{1s,.}\right)$.

\subsection{Sampling Idiosynchratic and Factor SV parameters \label{sec:Hamiltonian update}}

In the Particle Gibbs (PG) algorithm, each individual SV parameters
is drawn from the full conditional distribution $\mu_{1s}|h_{1s},\varTheta_{-\mu_{1s}}$,
$\phi_{1s}|h_{1s},\varTheta_{-\phi_{1s}}$, $\rho_{s}|h_{1s},\varTheta_{-\rho_{s}}$,
and $\tau_{1s}^{2}|h_{1s},\varTheta_{-\tau_{1s}^{2}}$, respectively.
For sampling $\tau_{1s}^{2}$, we obtain proposal from inverse gamma
distribution with $scale=\left(T-1\right)/2$ and $shape=M/2$, where
\[
M=\left(\left(1-\phi_{1s}^{2}\right)\left(h_{1s,1}-\mu_{1s}\right)^{2}+\sum_{t=2}^{T}\left(h_{1s,t}-\mu_{1s}-\phi_{1s}\left(h_{1s,t-1}-\mu_{1s}\right)\right)^{2}\right).
\]
Then, the acceptance probability is equal to $\min\left(1,R\right)$
with
\[
R=\frac{\prod_{t=1}^{T}p\left(y_{st}|h_{1st},h_{1st-1},\theta_{-\tau_{1s}^{2}},\tau_{1s}^{2*}\right)\left(1+\tau_{1s}^{2}\right)}{\prod_{t=1}^{T}p\left(y_{st}|h_{1st},h_{1st-1},\theta_{-\tau_{1s}^{2}},\tau_{1s}^{2}\right)\left(1+\tau_{1s}^{2*}\right)}.
\]
For sampling $\phi_{1s}$, we obtain proposal from $q\left(\phi_{1s}|h_{1s},\varTheta_{-\phi_{1s}}\right)\sim N\left(c_{\phi},d_{\phi}\right)$,
where
\[
d_{\phi}=\frac{\tau_{1s}^{2}}{\sum_{t=2}^{T}\left(h_{1s,t-1}-\mu_{1s}\right)^{2}-\left(h_{1s,1}-\mu_{1s}\right)^{2}}
\]
and
\[
c_{\phi}=d_{\phi}\frac{\sum_{t=2}^{T}\left(h_{1s,t}-\mu_{1s}\right)\left(h_{1s,t-1}-\mu_{1s}\right)}{\tau_{1s}^{2}}.
\]
The acceptance probability is equal to $\min\left(1,R\right)$ with
\[
R=\frac{\prod_{t=1}^{T}p\left(y_{st}|h_{1st},h_{1st-1},\theta_{-\phi_{1s}},\phi_{1s}^{*}\right)}{\prod_{t=1}^{T}p\left(y_{st}|h_{1st},h_{1st-1},\theta_{-\phi_{1s}},\phi_{1s}\right)}\sqrt{\frac{1+\phi_{1s}^{2*}}{1+\phi_{1s}^{2}}}.
\]
For sampling $\mu_{1s}$, we obtain proposal from normal distribution
$q\left(\mu_{1s}|h_{1s},\varTheta_{-\mu_{1s}}\right)\sim N\left(c_{\mu},d_{\mu}\right)$,
where
\[
d_{\mu}=\frac{\tau_{1s}^{2}}{1-\phi_{1s}^{2}+\left(T-1\right)\left(1-\phi_{1s}\right)^{2}}
\]
and
\[
c_{\mu}=d_{\mu}\frac{h_{1s,1}\left(1-\phi_{1s}^{2}\right)+\sum_{t=2}^{T}h_{1s,t}-\phi_{1s}h_{1s,t}+\phi_{1s}^{2}h_{1s,t-1}-\phi_{1s}h_{1s,t-1}}{\tau_{1s}^{2}}.
\]
The acceptance probability is equal to $\min\left(1,R\right)$ with
\[
R=\frac{\prod_{t=1}^{T}p\left(y_{st}|h_{1st},h_{1st-1},\theta_{-\mu_{1s}},\mu_{1s}^{*}\right)}{\prod_{t=1}^{T}p\left(y_{st}|h_{1st},h_{1st-1},\theta_{-\mu_{1s}},\mu_{1s}\right)}.
\]
The full conditional distribution can also be derived for $\mu_{2j}|h_{2j},\varTheta_{-\mu_{2j}}$,
$\phi_{2j}|h_{2j},\varTheta_{-\phi_{2j}}$, and $\tau_{2j}^{2}|h_{2j},\varTheta_{-\tau_{2j}^{2}}$.
For sampling $\tau_{2s}^{2}$, we obtain proposal from inverse gamma
distribution with $scale=\left(T-1\right)/2$ and $shape=M/2$, where
\[
M=\left(\left(1-\phi_{2s}^{2}\right)\left(h_{2s,1}-\mu_{2s}\right)^{2}+\sum_{t=2}^{T}\left(h_{2s,t}-\mu_{2s}-\phi_{2s}\left(h_{2s,t-1}-\mu_{2s}\right)\right)^{2}\right).
\]
Then, the acceptance probability is equal to $\min\left(1,R\right)$
with
\[
R=\frac{\left(1+\tau_{1s}^{2}\right)}{\left(1+\tau_{1s}^{2*}\right)}.
\]
For sampling $\phi_{2s}$, we obtain proposal from $q\left(\phi_{2s}|h_{2s},\varTheta_{-\phi_{2s}}\right)\sim N\left(c_{\phi},d_{\phi}\right)$,
where
\[
d_{\phi}=\frac{\tau_{2s}^{2}}{\sum_{t=2}^{T}\left(h_{2s,t-1}-\mu_{2s}\right)^{2}-\left(h_{2s,1}-\mu_{2s}\right)^{2}}
\]
and
\[
c_{\phi}=d_{\phi}\frac{\sum_{t=2}^{T}\left(h_{2s,t}-\mu_{2s}\right)\left(h_{2s,t-1}-\mu_{2s}\right)}{\tau_{2s}^{2}}.
\]
The acceptance probability is equal to $\min\left(1,R\right)$ with
\[
R=\sqrt{\frac{1+\phi_{1s}^{2*}}{1+\phi_{1s}^{2}}}.
\]
For sampling $\mu_{2s}$, we obtain from normal distribution $p\left(\mu_{2s}|h_{2s},\varTheta_{-\mu_{2s}}\right)\sim N\left(c_{\mu},d_{\mu}\right)$,
where
\[
d_{\mu}=\frac{\tau_{2s}^{2}}{1-\phi_{2s}^{2}+\left(T-1\right)\left(1-\phi_{2s}\right)^{2}}
\]
and
\[
c_{\mu}=d_{\mu}\frac{h_{2s,1}\left(1-\phi_{2s}^{2}\right)+\sum_{t=2}^{T}h_{2s,t}-\phi_{2s}h_{2s,t}+\phi_{2s}^{2}h_{2s,t-1}-\phi_{2s}h_{2s,t-1}}{\tau_{2s}^{2}}.
\]
Next, we discuss about the Hamiltonian Monte Carlo proposal to sample
the parameter $\rho_{s}$ for $s=1,...,p$ from conditional posterior
density $\tilde{\pi}\left(\rho_{s}|h_{1t},h_{2t},\varTheta_{-\rho_{s}}\right)$.
It can be used to generate distant proposals for the Particle Metropolis
within Gibbs algorithm to avoid the slow exploration behaviour that
results from simple random walk proposals. Suppose we want to sample
from a distribution with pdf proportional to $\exp\left(\mathcal{L}\left(\rho{}_{s}\right)\right)$,
where $\mathcal{L}\left(\rho{}_{s}\right)=\log\tilde{\pi}\left(\rho_{s}|h_{1t},h_{2t},\varTheta_{-\rho_{s}}\right)$
is the logarithm of the conditional posterior density of $\rho_{s}$
(up to a normalising constant). In Hamiltonian Monte Carlo \citep{Neal:2011},
we augment an auxiliary momentum variable $r_{s}$ for each parameter
$\rho{}_{s}$ with density $p\left(r_{s}\right)=N\left(r_{s}|0,1\right)$.
The joint density follows in factorised form as
\begin{eqnarray}
p\left(\rho{}_{s},r_{s}|h_{1t},h_{2t},\varTheta_{-\rho_{s}},\boldsymbol{y}\right) & \propto & \exp\left(\mathcal{L}\left(\rho{}_{s}\right)-\frac{1}{2}r_{s}^{2}\right)\nonumber \\
 & \propto & \exp\left(-H\left(\rho{}_{s},r_{s}\right)\right).\label{eq:jointHamiltonian-1}
\end{eqnarray}
This augmented model can be interpreted as Hamiltonian system where
$\rho{}_{s}$ denotes a parameter's position, $r_{s}$ denotes the
momentum, $\mathcal{L}\left(\rho{}_{s}\right)$ is a negative potential
energy function of the parameters $\rho_{s}$, and $\frac{1}{2}r_{s}^{2}$
is the kinetic energy function of the parameters, and $-H\left(\rho{}_{s},r_{s}\right)$
is the total negative energy of the parameters and momentum variables
and the function $H\left(\rho{}_{s},r_{s}\right)$ is often called
Hamiltonian. At the end of this algorithm, we will discard the momentum
variable $r_{s}$, obtaining a new $\rho{}_{s}$ that is still distributed
as $\exp\left(\mathcal{L}\left(\rho{}_{s}\right)\right)$. Equation
\eqref{eq:jointHamiltonian-1} is factorisable because the conditional
distribution of momentum does not depend on the parameter values.

In the Hamiltonian dynamics, the parameters $\rho_{j}$ and the momentum
variables $r_{j}$ are moved along a continuous time $\boldsymbol{t}$
according to the following differential equations
\[
\frac{d\rho_{s}}{dt}=\frac{\partial H}{\partial r_{s}}=r_{s}
\]

\[
\frac{dr_{s}}{dt}=-\frac{\partial H}{\partial\rho_{s}}=\nabla_{\rho_{s}}\mathcal{L}\left(\rho{}_{s}\right),
\]
where $\nabla_{\rho_{s}}$ denotes the gradient with respect to the
parameter $\rho_{s}$. In implementation, this Hamiltonian dynamics
needs to be approximated by discretised time, using small step size
$\epsilon$. We can simulate the evolution over time of $\left(\rho{}_{s},r_{s}\right)$
via ``leapfrog'' integrator. The one step Leapfrog update is given
as
\begin{eqnarray*}
r_{s}\left(t+\frac{\epsilon}{2}\right) & = & r_{s}\left(t\right)+\epsilon\nabla_{\rho{}_{s}}\mathcal{L}\left(\rho_{s}\left(t\right)\right)/2\\
\rho_{s}\left(t+\epsilon\right) & = & \rho_{s}\left(t\right)+\epsilon r_{s}\left(t+\frac{\epsilon}{2}\right)\\
r_{s}\left(t+\epsilon\right) & = & r_{s}\left(t+\epsilon/2\right)+\epsilon\nabla_{\rho{}_{s}}\mathcal{L}\left(\rho_{s}\left(t+\epsilon\right)\right)/2
\end{eqnarray*}
Each leapfrog step is time reversible by negation of the step size,
$\epsilon$. Since leapfrog integrator provides mapping $\left(\rho{}_{s},r_{s}\right)\rightarrow\left(\rho{}_{s}^{*},r_{s}^{*}\right)$
that are both time-reversible and volume preserving \citep{Neal:2011},
then the Metropolis-Hastings algorithm with acceptance probability
given by $\min\left(1,\frac{\exp\left(\mathcal{L}\left(\rho_{s}^{*}\right)-\frac{1}{2}r_{s}^{2*}\right)}{\exp\left(\mathcal{L}\left(\rho{}_{s}\right)-\frac{1}{2}r_{s}^{2}\right)}\right)$
produces an ergodic, time reversible Markov chain that satisfies detailed
balance and whose stationary density is $p\left(\rho{}_{s},r_{s}|h_{1t},h_{2t},\varTheta_{-\rho_{s}},\boldsymbol{y}\right)$
\citep{Liu:2001a,Neal:1996}. A summary of the Hamiltonian Monte Carlo
algorithm is given in Algorithm \ref{alg:Hamiltonian-Monte-Carlo}.

\begin{algorithm}[H]
\caption{Hamiltonian Monte Carlo \label{alg:Hamiltonian-Monte-Carlo}}

Given $\rho{}_{s}^{0}$, $\epsilon$, $Leap$, $S$, where $Leap$
is the number of Leapfrog updates.
\begin{itemize}
\item For $l=1$ to $L$

Sample $r_{s}^{0}\sim N\left(0,1\right)$.

Set $\rho{}_{s}^{l}\leftarrow\rho_{s}^{l-1}$, $\rho{}_{s}^{*}\leftarrow\rho_{s}^{l-1}$,
and $r_{s}^{*}\leftarrow r_{s}^{0}$.
\begin{itemize}
\item For $i=1$ to $Leap$

Set $\left(\rho{}_{s}^{*},r_{s}^{*}\right)\leftarrow Leapfrog\left(\rho{}_{s}^{*},r_{s}^{*},\epsilon\right)$

end for

\end{itemize}

With probability $\alpha=\min\left(1,\frac{\exp\left(\mathcal{L}\left(\rho_{s}^{*}\right)-\frac{1}{2}r_{s}^{2*}\right)}{\exp\left(\mathcal{L}\left(\rho{}_{s}\right)-\frac{1}{2}r_{s}^{2}\right)}\right)$,
set $\rho{}_{s}^{l}=\rho{}_{s}^{*}$, $r_{s}^{l}=-r_{s}^{*}$.

end for\end{itemize}
\end{algorithm}

The performance of HMC depends strongly on choosing suitable values
for $\epsilon$ and $L$. The step size $\epsilon$ determines how
well the leapfrog integration can approximate Hamiltonian dynamics.
If we set $\epsilon$ too large, then the simulation error is large
and yield low acceptance rate. However, if we set $\epsilon$ too
small, then the computational burden is high to obtain distant proposals.
In the same way, if we set $L$ too small, the proposal will be close
to the current value of parameters, resulting in undesirable random
walk behaviour and slow mixing. If $L$ is too large, HMC will generate
trajectories that retrace back their steps. In this paper, we use
No-U-Turn sampler (NUTS) with dual averaging algorithm developed by
\citet{Hoffman:2014} and \citet{Nesterov:2009}, respectively, that
still leaves the target density invariant and satisfies time reversibility
to adaptively select $L$ and $\epsilon$, respectively.

In the Mixed sampler, for sampling $\tau_{1s}^{2}$ for $s=1,...,p$
and $\tau_{2j}^{2}$ for $j=1,...,k$ are done in Pseudo Marginal
(PM) step. In PM step, the gradient of log-posterior cannot be computed
exactly and need to be estimated. The efficiency of PMMH will then
depend crucially on how accurately we can estimate the gradient of
log-posterior. If the error in the estimate of the gradient is too
large, then there will be no advantage in using proposals with derivatives
information over a random walk proposal \citep{Nemeth:2016}. In this
paper, we employ a single step of the leapfrog algorithm that has
an update of the form
\begin{eqnarray*}
\tau_{j}^{2}\left(t+\epsilon\right) & = & \tau_{j}^{2}\left(t\right)+\frac{\epsilon^{2}}{2}\nabla_{\tau_{j}^{2}}\mathcal{L}_{\tau_{j}^{2}}\left(\tau_{j}^{2}\left(t\right)\right)+\epsilon r_{j}\left(t\right)\\
r_{j}\left(t+\epsilon\right) & = & r_{j}\left(t\right)+\frac{\epsilon}{2}\nabla_{\tau_{j}^{2}}\mathcal{L}_{\tau_{j}^{2}}\left(\tau_{j}^{2}\left(t\right)\right)+\frac{\epsilon}{2}\nabla_{\tau_{j}^{2}}\mathcal{L}_{\tau_{j}^{2}}\left(\tau_{j}^{2}\left(t+\epsilon\right)\right).
\end{eqnarray*}
This update is a discrete pre-conditioned Langevin diffusion as employed
in Metropolis Adjusted Langevin Algorithm (MALA) \citep{Roberts:2003}.The
algorithm to estimate the gradient of log-posterior is given in Appendix
\ref{alg:Algorithm-to-estimate gradient and hessian}.

\section{Simulation Study\label{sec:Simulation-Study}}

In order to compare different sampling schemes in terms of sampling
efficiency, a simple simulation study is conducted. We use $T=1000$
periods of data using a model with $k=2$ factors and $p=10$ dimensions.
We set $\phi_{1s}=0.98$, $\rho_{1s}=-0.1$, $\mu_{1s}=0.01$, and
$\tau_{1s}^{2}=0.05$, for all $s=1,...,p$, and also $\phi_{2j}=0.98$,
and $\tau_{2j}^{2}=0.05$ for $j=1,...,k$, and
\[
B^{'}=\left[\begin{array}{cccccccccc}
1 & 0.9 & 0.8 & 0.7 & 0.6 & 0.5 & 0.4 & 0.3 & 0.2 & 0.1\\
0 & 1 & 0.1 & 0.2 & 0.3 & 0.4 & 0.5 & 0.6 & 0.7 & 0.8
\end{array}\right]
\]
In this simulation study, the total number of MCMC iterations is $15000$,
with the first $5000$ discarded as burn in replications. The number
of particles is $500$. We conduct a simulation study in order to
compare three different approaches to estimation: PG, particle Gibbs
with additional ancestor sampling step (PGAS), and Mixed samplers.
To define our measure of the inefficiency of different sampling schemes
that takes computing time into account, we first define the Integrated
Autocorrelation Time $\left(IACT_{\boldsymbol{\theta}}\right)$. For
a univariate parameter $\theta$, IACT is estimated by
\[
IACT\left(\theta_{1:M}\right)\coloneqq1+2\sum_{t=1}^{L}\widehat{\rho}_{t}\left(\theta_{1:M}\right),
\]
where $\widehat{\rho}_{t}\left(\theta_{1:M}\right)$ denotes the empirical
autocorrelation at lag $t$ of $\theta_{1:M}$ (after the burnin periods
have been discarded). A lower value of IACT indicates that the chain
mixed well. Here, $L$ is chosen as the first index for which the
empirical autocorrelation satisfies $\left|\widehat{\rho}_{t}\left(\theta_{1:M}\right)\right|<2/\sqrt{M}$,
i.e. when the empirical autocorrelation coefficient is statistically
insignificant. Our measure of inefficiency of sampling scheme is the
time normalised variance
\begin{equation}
TNV\coloneqq IACT_{mean}\times CT,\label{eq:TNV}
\end{equation}
where $CT$ is the computing time and $IACT_{mean}$ be the mean of
IACT's over all parameters.

\begin{figure}[H]
\caption{Plots of estimated trajectories against the true trajectories of idiosynchratic
log-variances $h_{1s,t}$ for $s=1,...,10$ \label{fig:Plots-of-estimated trajectory simulation idiosynchratic}}

\centering{}\includegraphics[width=15cm,height=10cm]{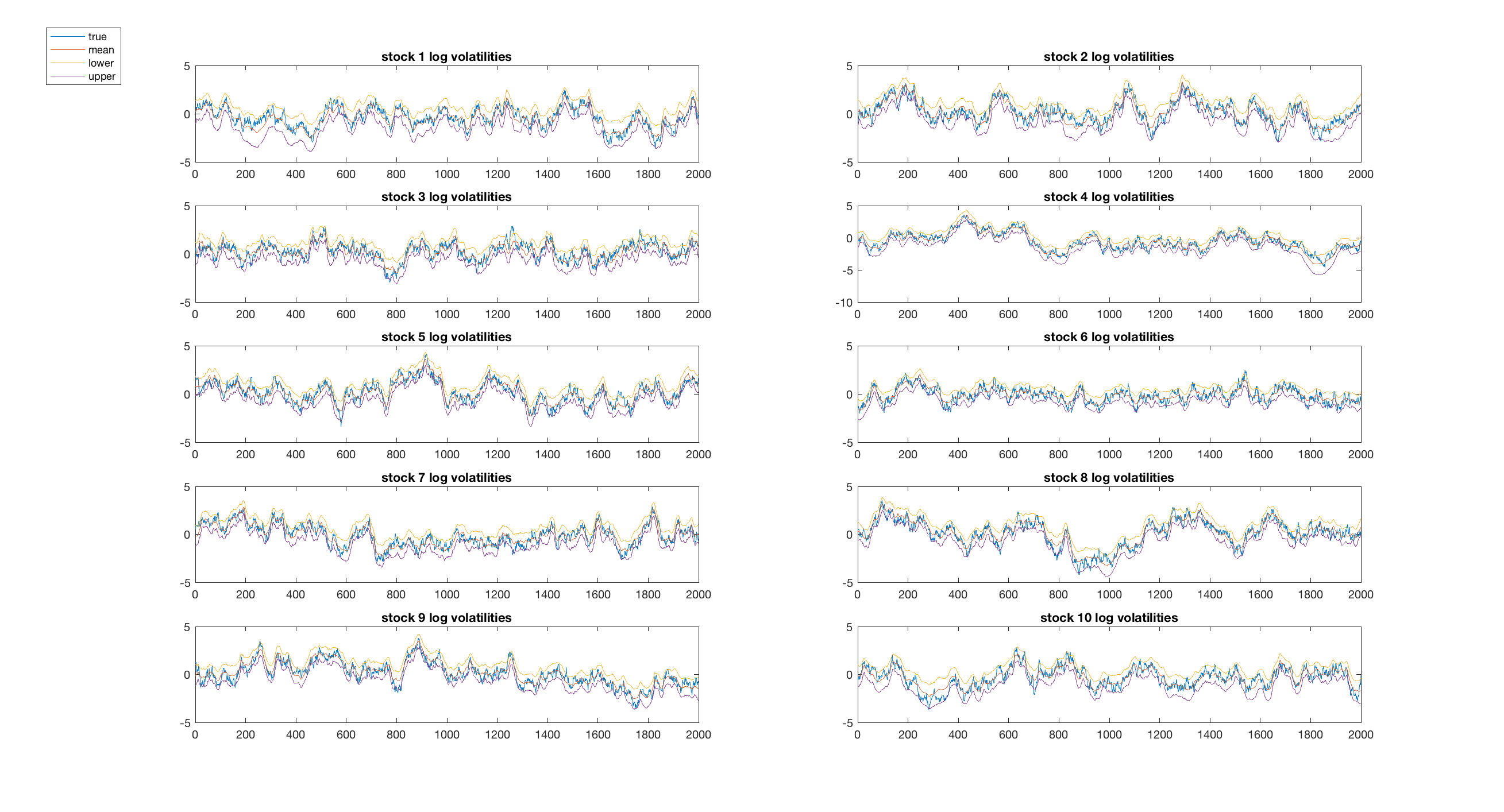}
\end{figure}

\begin{figure}[H]
\caption{Plot of estimated trajectories against the true trajectories of the
factor log-variances $h_{2j,t}$ for $j=1,...,2$\label{fig:Plot-of-estimated trajectory factor}}

\centering{}\includegraphics[width=15cm,height=10cm]{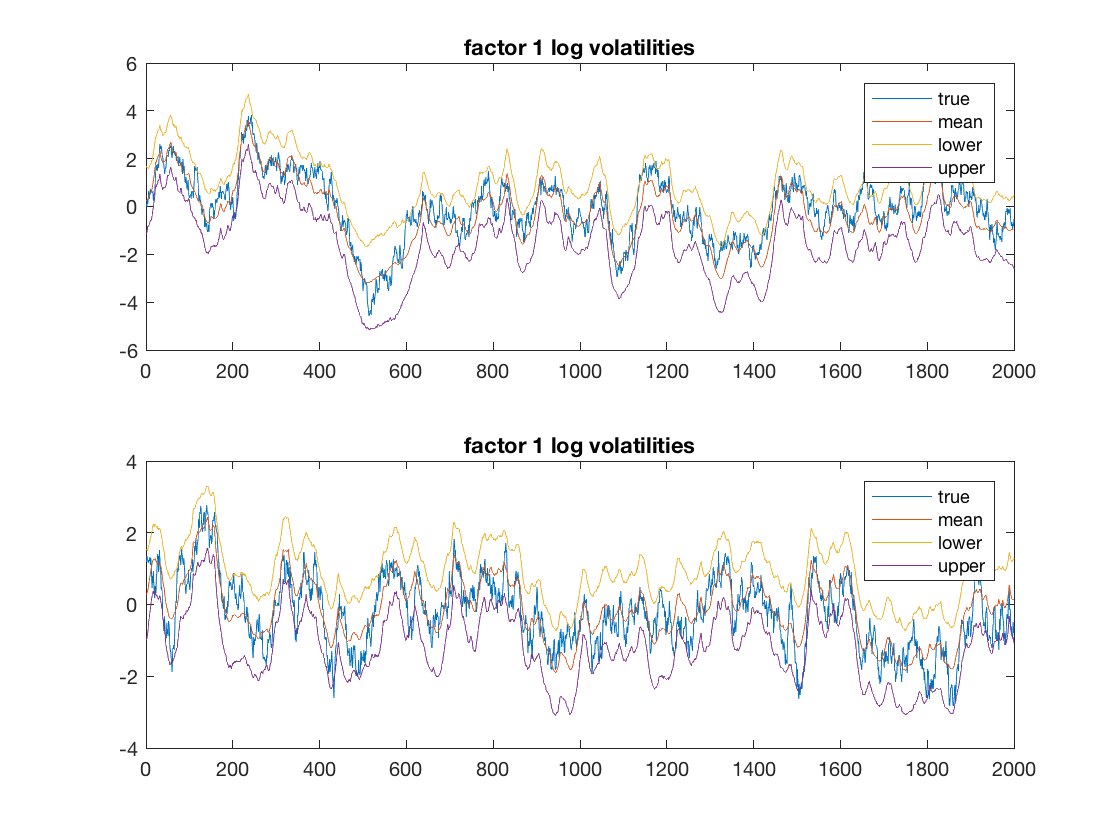}
\end{figure}

Tables \ref{tab:Multivariate-Factor-SV sim result mixed}, \ref{tab:Multivariate-Factor-SV sim result PG},
and \ref{tab:Multivariate-Factor-SV sim result PGAS} in the Appendix
show the inefficiency factors for all the parameters of factor MSV
with leverage for the mixed, PG, and PGAS samplers, respectively.
Table \ref{tab:Comparison-of-Different sampling scheme simulation}
summarises the simulation results and shows that the mixed sampler
is more than $4$ times more efficient than the PG sampler and more
than 2 times more efficient than PGAS sampler. PGAS sampler is 2 times
more efficient than PG sampler. Figures \ref{fig:Plots-of-estimated trajectory simulation idiosynchratic}
and \ref{fig:Plot-of-estimated trajectory factor} show the estimated
trajectory of idiosynchratic log-variances $h_{1s,t}$ for $s=1,...,p$
and factor log variances $h_{2j,t}$ for $j=1,..,k$ from mixed sampler.
They estimate the true trajectory of idiosynchratic and factor log
variances well.

\begin{table}[H]
\caption{Comparison of Different Sampling Schemes. Time in seconds ($N=500$)
PG: Particle Gibbs, PGAS: Particle Gibbs with additional ancestor
sampling step, Mix.: $PMMH\left(\tau_{11}^{2},...,\tau_{1p}^{2},\tau_{21}^{2},...,\tau_{2k}^{2}\right)+PG\left(\mu_{1s},\phi_{1s},\rho_{1s},s=1,...,p;\mu_{1j},\phi_{1j},j=1,...,k;B,f\right)$.
\label{tab:Comparison-of-Different sampling scheme simulation}}

\centering{}%
\begin{tabular}{cccc}
\hline
 & PG & PGAS & Mix.\tabularnewline
\hline
Time & 1.07 & 1.18 & 1.75\tabularnewline
$IACT_{Mean}$ & 70.45 & 31.05 & 10.42\tabularnewline
TNV & 75.38 & 36.64 & 18.23\tabularnewline
Rel. TNV & 4.13 & 2.01 & 1\tabularnewline
\hline
\end{tabular}
\end{table}

\section{Empirical Application to US stock returns\label{sec:Empirical-Application-to}}

We applied the estimation described above to a sample of daily US
stock returns. The data, provided by Kenneth French, consisted of
the daily returns for 16 industry portfolios and it is given in Table
\ref{tab:List-of-Industry}. We used a sample running from 13rd August
1993 to 17th July 2001, a total of 2000 observations. We consider
models with $1-4$ factors. The total number of MCMC iterations is
$15000$, with the first $5000$ discarded as burn in replications.
The number of particles is $1000$. Table \ref{tab:Comparison-of-Different sampling schemes real application}
summarises the estimation results for $1-4$ factors with standard
normal prior $N\left(0,1\right)$ for each elements of factor loading
matrix. It is clear that mixed sampler is always better than the PGAS
sampler for all cases. Given the outcome of these comparisons, the
remaining analysis are based on the results from the mixed sampler.

\begin{table}[H]
\begin{centering}
\caption{List of Industry Portfolios\label{tab:List-of-Industry}}
\begin{tabular}{cccccccc}
\hline
 & Industry &  & Industry &  & Industry &  & Industry\tabularnewline
\hline
\hline
1 & Automobiles & 5 & Fabricated & 9 & Mining & 13 & Steel\tabularnewline
2 & Chemicals & 6 & Banks & 10 & Oil & 14 & Transportation\tabularnewline
3 & Textiles & 7 & Food & 11 & Other & 15 & Consumer Durables\tabularnewline
4 & Drugs, Soap, etc. & 8 & Machinery & 12 & Retail & 16 & Utilities\tabularnewline
\hline
\end{tabular}
\par\end{centering}

\end{table}

\begin{table}[H]
\caption{Comparison of Different Sampling Schemes for Factor MSV with leverage.
Time in seconds ($N=1000$) \label{tab:Comparison-of-Different sampling schemes real application}}

\centering{}%
\begin{tabular}{ccccccccc}
\hline
 & \multicolumn{2}{c}{1 Factor} & \multicolumn{2}{c}{2 Factor} & \multicolumn{2}{c}{3 factor} & \multicolumn{2}{c}{4 factor}\tabularnewline
\hline
 & PGAS & Mixed & PGAS & Mixed & PGAS & Mixed & PGAS & Mixed\tabularnewline
\hline
Time & 2.63 & 4.35 & 2.65 & 4.40 & 2.70 & 4.45 & 2.75 & 4.50\tabularnewline
$IACT_{Mean}$ & 132.04 & 15.96 & 122.04 & 16.69 & 111.42 & 29.76 & 146.11 & 26.01\tabularnewline
TNV & 347.27 & 69.43 & 323.41 & 73.44 & 300.83 & 132.43 & 401.80 & 117.05\tabularnewline
Rel. TNV & 5.00 & 1 & 4.40 & 1 & 2.27 & 1 & 3.43 & 1\tabularnewline
\hline
\end{tabular}
\end{table}

In this paper, we select the number of factor using deviance information
criterion (DIC). In the seminal paper, \citet{Spiegelhalter:2002}
proposed a concept of deviance information criterion (DIC) for model
comparison. The model selection is based on the deviance, which is
given by
\[
D\left(\theta\right)=-2\log p\left(y|\varTheta\right)+2\log h\left(y\right),
\]
where $p\left(y|\varTheta\right)$ is the likelihood function of the
parametric model and $h\left(y\right)$ is some fully specified standardising
term that is only a function of the data. For model comparison purposes,
we set $h\left(y\right)=1$ for all models. The effective number of
parameters $p_{D}$ is defined as
\[
p_{D}=\overline{D\left(\theta\right)}-D\left(\tilde{\theta}\right),
\]
where
\[
\overline{D\left(\theta\right)}=-2E_{\theta}\left[\log p\left(y|\varTheta\right)|y\right]+2\log h\left(y\right)
\]
is the posterior mean deviance and $\tilde{\theta}$ is an estimate
of $\theta$, which is usually set to be posterior mode or mean. Thus,
the deviance information criterion is defined as
\begin{eqnarray*}
DIC & = & \overline{D\left(\theta\right)}+p_{D}\\
 & = & -4E_{\theta}\left[\log p\left(y|\varTheta\right)|y\right]+2\log p\left(y|\tilde{\theta}\right).
\end{eqnarray*}
Given a set of models for the given data, the preferred model is the
one with the minimum DIC value. \citet{Celeux:2006} pointed that
there are a number of alternative definitions of the DIC in the latent
variable models. In this paper, we follow the definitions of DIC that
are based on conditional likelihood, which is given by:

\[
DIC_{7}=-4E_{\theta,Z}\left[\log p\left(y|\varTheta,Z\right)|y\right]+2\log p\left(y|\widehat{Z},\widehat{\theta}\right),
\]
where $\left(\widehat{Z},\widehat{\theta}\right)$ is the joint maximum
a posterior (MAP) estimate, and $Z$ consists of all the latent volatilities
in the model. The first term on the right hand side can be estimated
by averaging the log-conditional likelihoods $\log p\left(y|\varTheta,Z\right)$
over the posterior draws of $\left(Z,\theta\right)$.

\begin{figure}[H]
\caption{Marginal Posteriors of the factor log-variances $(mean\pm2sd)$, $h_{2j,t}$
for $j=1,2,3,4$\label{fig:Marginal-Posteriors-of factor log variances}}

\centering{}\includegraphics[width=15cm,height=10cm]{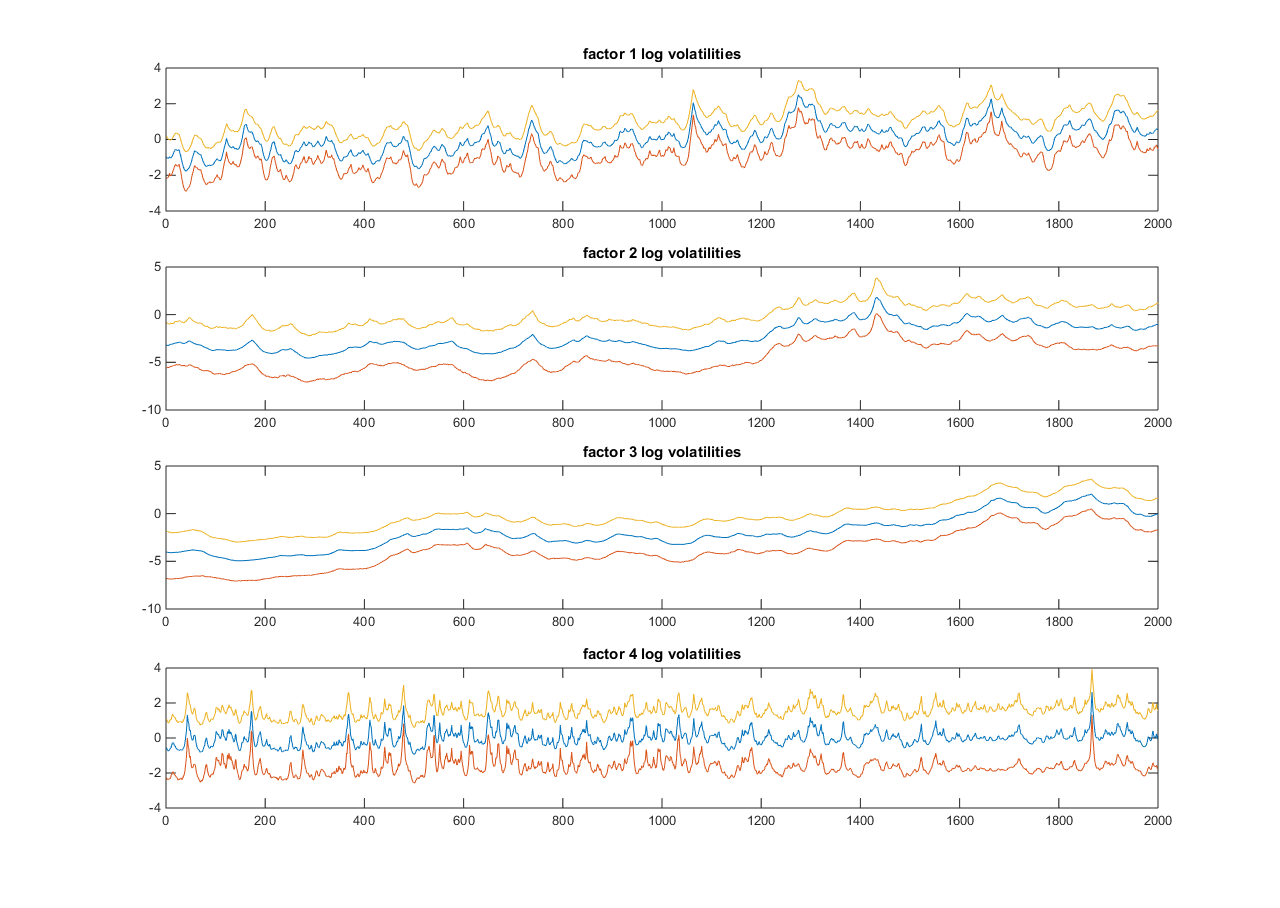}
\end{figure}

\begin{figure}[H]
\caption{Posterior Volatilities of US Log returns $(mean\pm2sd)$, $h_{1s,t}$
for $s=1,...,16$\label{fig:Posterior-Volatilities-of log returns}}

\centering{}\includegraphics[width=15cm,height=10cm]{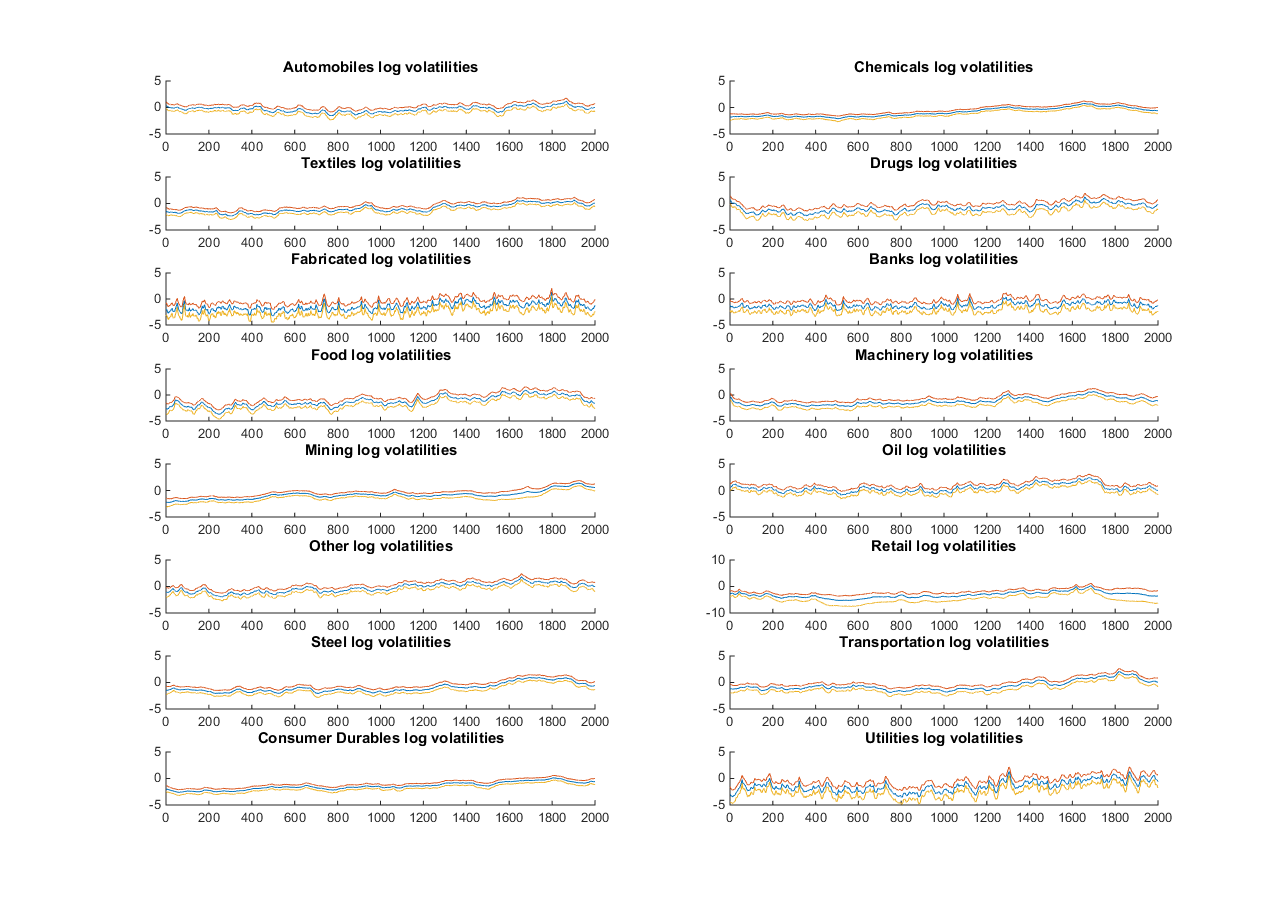}
\end{figure}

Table \ref{tab:Selecting-Number-of factor} shows the DIC values for
1-4 factors with different priors for elements of factor loading matrix.
We compare standard normal prior $N\left(0,1\right)$, normal-gamma
prior with $a_{s}=1$ and $a_{s}=0.5$ for all $s=1,...,p$. We set
the hyper-hyperparameters $c_{s}=d_{s}=2$ for all $s=1,...,p$. The
best model is the four-factor model with normal-gamma prior $\left(a_{s}=0.5\right)$.
{\em We found no evidence for the leverage effects in
the dataset, with posterior credible intervals of each $\rho_{s}$
including zero}, except the mining industry. We begin by discussing
the log-variances of the latent factors, visualised in Figure \ref{fig:Marginal-Posteriors-of factor log variances}
and the corresponding posterior means of the factor loadings given
in Table \ref{tab:Posterior-Means-of factor loading matrix}. The
first factor can clearly be interpreted as the mining industry driven
one. The automobiles, transportation and steel industries also load
very highly on this factor. Factor 1's log-variance appears quite
volatile throughtout the sample period. Factor 2's log variances appears
slightly less volatile than the first factor, and generally very smooth
and more persistent. It is also driven by mining industry. Retail
and steel industries also loads very highly on this factor. The third
factor volatility shows a similar overall pattern as the second. The
fourth factor volatility shows a similar pattern as the first, but
slightly more volatile. Figure \ref{fig:Posterior-Volatilities-of log returns}
shows the marginal posterior means of univariate volatilities for
all $16$ US stock returns from 13rd August 1993 to 17th July 2001.
In general, the log-volatilities are generally smooth and less volatile,
except, the fabricated, utilities, and bank industries.

\begin{table}[H]
\caption{Selecting Number of Factors based on mixed sampler using DIC criterion
\label{tab:Selecting-Number-of factor}}

\centering{}%
\begin{tabular}{cccc}
\hline
Number of Factors & $N\left(0,1\right)$ prior & N-G prior $\left(a_{s}=1\right)$ & N-G prior $\left(a_{s}=0.5\right)$\tabularnewline
\hline
1 & $73893.38$ & $73965.72$ & NA\tabularnewline
2 & $72607.25$ & $72466.73$ & NA\tabularnewline
3 & $71594.45$ & $71626.78$ & $71490.18$\tabularnewline
4 & $71288.95$ & $71334.78$ & $71285.17$\tabularnewline
\hline
\end{tabular}
\end{table}

\begin{table}[H]
\caption{The Inefficiency of factor loading matrix for different priors\label{tab:Selecting-Number-of factor-1}}

\centering{}%
\begin{tabular}{cccc}
\hline
Number of Factors & $N\left(0,1\right)$ prior & N-G prior $\left(a_{s}=1\right)$ & N-G prior $\left(a_{s}=0.5\right)$\tabularnewline
\hline
3 & $18.65$ & $29.56$ & $22.46$\tabularnewline
4 & $26.77$ & $27.01$ & $28.95$\tabularnewline
\hline
\end{tabular}
\end{table}

\begin{table}[H]
\caption{Posterior Means of factor loading matrix of four factor models with
N-G prior $\left(a_{s}=0.5\right)$ \label{tab:Posterior-Means-of factor loading matrix}}

\centering{}%
\begin{tabular}{ccccc}
\hline
Automobiles & 0.73 & 0 & 0 & 0\tabularnewline
Chemicals & 0.68 & -0.59 & 0 & 0\tabularnewline
Textiles & 0.58 & 0.05  & 0.21 & 0\tabularnewline
Drugs, Soap, etc. & 0.58 & 0.48 & -0.26 & 0.42\tabularnewline
Fabricated & 0.62 & -0.01  & 0.21 & 0.03\tabularnewline
Banks & 0.56  & -0.50 & 0.24 & 0.02\tabularnewline
Food & 0.71  & 0.21 & 0.17  & 0.19\tabularnewline
Machinery & 0.51 & 0.10 & -0.32 & 0.38\tabularnewline
Mining & 0.83  & 0.71 & 1.61 & 0.11\tabularnewline
Oil & 0.39  & -0.05 & 0.11 & -0.11\tabularnewline
Other & 0.48  & -0.59  & -0.12 & 0.17\tabularnewline
Retail & 0.70 & 0.57 & 0.94 & 0.17\tabularnewline
Steel & 0.72  & 0.61 & 0.01 & 0.09 \tabularnewline
Transportation & 0.72  & -0.67 & 0.85 & -0.10\tabularnewline
Consumer Durables & 0.70  & -0.25 & 0.11  & 0.03\tabularnewline
Utilities & 0.33 & -0.15 & -0.16 & 0.21\tabularnewline
\hline
\end{tabular}
\end{table}

\section{Conclusions\label{sec:Conclusions}}

Estimating time-varying covariance matrices of financial times series
is an active area of research. In this paper, we employ factor multivariate
stochastic volatility (factor MSV) models with leverage because they
are able to model the volatility dynamics of a large system of financial
or economic time series when the common features in these series can
be captured by a small number of latent factors. To conduct efficient
and reliable statistical inference, we propose a sampler based on
recent developments in PMCMC methods. Our article demonstrates that
a version of general PMCMC sampler of \citet{Mendes2016} provides
a flexible and efficient framework to carry out inference on factor
MSV models with leverage. The resulting parameter estimates mix well.
The proposed method is illustrated using simulated and real datasets.

\appendix

\section{Generic Sequential Monte Carlo (SMC) Algorithm \label{sec:Generic-Sequential-Monte}}
\begin{enumerate}
\item For $t=1$

\begin{enumerate}
\item Sample $h_{1}^{i}$ from $m_{1}^{\theta}\left(h\right)$, for $i=1,...,N$
\item Calculate the importance weights
\[
w_{1}^{i}=\frac{p_{\theta}\left(y_{1}|h_{1}^{i}\right)p_{\theta}\left(h_{1}^{i}\right)}{m_{1}^{\theta}\left(h_{1}^{i}\right)},i=1,...,N.
\]
and normalised them to obtain $\overline{w}_{1}^{1:N}$.
\end{enumerate}
\item For $t>1$

\begin{enumerate}
\item Sample the ancestral indices $C_{t-1}^{1:N}\sim M\left(a^{1:N}|\overline{w}_{t-1}^{1:N}\right)$.
\item Sample $h_{t}^{i}$ from $m_{t}^{\theta}\left(h_{t}|h_{t-1}^{a_{t-1}^{i}}\right)$,
$i=1,...,N$.
\item Set $h_{1:t}^{\left(i\right)}=\left(h_{1:t-1}^{\left(a_{t-1}^{i}\right)},h_{t}^{\left(i\right)}\right)$.
\item Calculate the importance weights
\[
w_{t}^{i}=\frac{p_{\theta}\left(y_{1}|h_{t}^{i}\right)p_{\theta}\left(h_{t}^{i}|h_{t-1}^{a_{t-1}^{i}}\right)}{m_{t}^{\theta}\left(h_{t}^{i}|h_{t-1}^{a_{t-1}^{i}}\right)},i=1,...,N.
\]
and normalised them to obtain $\overline{w}_{t}^{1:N}$.
\end{enumerate}
\end{enumerate}

\section{Conditional Sequential Monte Carlo\label{sec:Conditional-Sequential-Monte}}

In the following, we describe the steps of conditional particle filter
to draw $\left(h_{1:T}\right)$ \citep{Andrieu:2010}.
\begin{enumerate}
\item Fix $h_{1:T}^{j}$ and $C_{1:T-1}^{j}$.
\item For $t=1$

\begin{enumerate}
\item Sample $h_{1}^{i}$ from $m_{1}^{\theta}\left(h_{1}\right)$, for
$i\in\left\{ 1,...,N\right\} \setminus\left\{ b_{1}^{j}\right\} $.
\item Calculate the importance weights
\[
w_{1}^{i}=\frac{p_{\theta}\left(y_{1}|h_{1}^{i}\right)p_{\theta}\left(h_{1}^{i}\right)}{m_{1}^{\theta}\left(h_{1}^{i}\right)},i=1,...,N.
\]
and normalised them to obtain $\overline{w}_{1}^{1:N}$.
\end{enumerate}
\item For $t>1$

\begin{enumerate}
\item Sample the ancestral indices $C_{t-1}^{-\left(b_{t}^{j}\right)}\sim M\left(a^{\left(-b_{t}^{j}\right)}|\overline{w}_{t-1}^{1:N}\right)$.
\item Sample $h_{t}^{i}$ from $m_{t}^{\theta}\left(h_{t}|h_{t-1}^{a_{t-1}^{i}}\right)$,
$i=1,...,N\setminus\left\{ b_{t}^{j}\right\} $.
\item Set $h_{1:t}^{\left(i\right)}=\left(h_{1:t-1}^{\left(a_{t-1}^{i}\right)},h_{t}^{\left(i\right)}\right)$,
$i=1,...,N$.
\item Calculate the importance weights
\[
w_{t}^{i}=\frac{p_{\theta}\left(y_{1}|h_{t}^{i}\right)p_{\theta}\left(h_{t}^{i}|h_{t-1}^{a_{t-1}^{i}}\right)}{m_{t}^{\theta}\left(h_{t}^{i}|h_{t-1}^{a_{t-1}^{i}}\right)},i=1,...,N.
\]
and normalised them to obtain $\overline{w}_{t}^{1:N}$.
\end{enumerate}
\end{enumerate}

\section{Conditional Sequential Monte Carlo for Ancestral Sampling Algorithm\label{sec:Conditional-Sequential-Monte with Ancestral Sampling}}

In the following, we describe the steps of conditional particle filter
with ancestor sampling to draw $\left(h_{1:T}\right)$ \citep{Lindsten:2014}.
\begin{enumerate}
\item Fix $h_{1:T}^{j}$ and $C_{1:T-1}^{j}$.
\item For $t=1$

\begin{enumerate}
\item Sample $h_{1}^{i}$ from $m_{1}^{\theta}\left(h_{1}\right)$, for
$i\in\left\{ ,...,N\right\} \setminus\left\{ b_{1}^{j}\right\} $.
\item Calculate the importance weights
\[
w_{1}^{i}=\frac{p_{\theta}\left(y_{1}|h_{1}^{i}\right)p_{\theta}\left(h_{1}^{i}\right)}{m_{1}^{\theta}\left(h_{1}^{i}\right)},i=1,...,N.
\]
and normalised them to obtain $\overline{w}_{1}^{1:N}$.
\end{enumerate}
\item For $t>1$

\begin{enumerate}
\item Sample the ancestral indices $C_{t-1}^{-\left(b_{t}^{j}\right)}\sim M\left(a^{\left(-b_{t}^{j}\right)}|\overline{w}_{t-1}^{1:N}\right)$.
\item Draw $b_{t-1}^{\left(j\right)}$ from $p\left(b_{t-1}^{\left(j\right)}=k\right)\propto w_{t-1}^{\left(k\right)}p_{\theta}\left(h_{t}^{b_{t}^{\left(j\right)}}|h_{t-1}^{\left(k\right)}\right)$.
\item Sample $h_{t}^{i}$ from $m_{t}^{\theta}\left(h_{t}|h_{t-1}^{a_{t-1}^{i}}\right)$,
$i=1,...,N\setminus\left\{ b_{t}^{j}\right\} $.
\item Calculate the importance weights
\[
w_{t}^{i}=\frac{p_{\theta}\left(y_{1}|h_{t}^{i}\right)p_{\theta}\left(h_{t}^{i}|h_{t-1}^{a_{t-1}^{i}}\right)}{m_{t}^{\theta}\left(h_{t}^{i}|h_{t-1}^{a_{t-1}^{i}}\right)},i=1,...,N.
\]
and normalised them to obtain $\overline{w}_{t}^{1:N}$.
\end{enumerate}
\end{enumerate}

\section{Estimating Gradients of Log-Posterior using Particle Filter\label{sec:Estimating-Gradients-of}}

This section presents the construction of the proposal density in
Pseudo Marginal (PM) step that makes use of the derivatives of the
log likelihood. \citet{Poyiadjis:2011} were the first to show how
the particle filter methods can be used to estimate the derivatives
of the log likelihood for state space models. Their methods might
suffer from a computational cost that is quadratic in the number of
particles, \citet{Nemeth:2016} proposed an alternative method whose
computational cost is linear in the number of particles. They use
a combination of kernel density estimation and Rao-Blackwellisation
to reduce the Monte Carlo error of the estimates.

For non-linear and non-Gaussian state space models it is not possible
to obtain the score and observed information matrix exactly. If it
is possible to obtain a particle approximation of $p\left(\boldsymbol{h}_{1:T}|\boldsymbol{y}_{1:T},\theta\right)$,
then this approximation can be used to estimate the score vector $\nabla\log p\left(\boldsymbol{y}_{1:T}|\theta\right)$
using Fisher's identity \citep{Cappe:2005}
\[
\nabla\log p\left(\boldsymbol{y}_{1:T}|\theta\right)=\int\nabla\log p\left(\boldsymbol{h}_{1:T},\boldsymbol{y}_{1:T}|\theta\right)p\left(\boldsymbol{h}_{1:T}|\boldsymbol{y}_{1:T},\theta\right)d\boldsymbol{h}_{1:T}.
\]
where
\[
\nabla\log p\left(\boldsymbol{h}_{1:T},\boldsymbol{y}_{1:T}|\theta\right)=\sum_{t=1}^{T}\left\{ \nabla\log g_{\theta}\left(\boldsymbol{y}_{t}|\boldsymbol{h}_{t}\right)+\nabla\log p_{\theta}\left(\boldsymbol{h}_{t}|\boldsymbol{h}_{t-1}\right)\right\} .
\]
The algorithm to estimate Gradient is given in \eqref{alg:Algorithm-to-estimate gradient and hessian}.

\begin{algorithm}[H]
\caption{Algorithm to estimate Gradient and Hessian Matrix \citet{Nemeth:2016}
\label{alg:Algorithm-to-estimate gradient and hessian}}

\begin{itemize}
\item Initialise: set $m_{0}^{\left(i\right)}=0$ and $n_{0}^{\left(i\right)}=0$
for $i=1,...,N$, where $N$ is the number of particles, and $S_{0}=0$
and $B_{0}=0$.
\item At iteration $t=1,...,T$

\begin{itemize}
\item Run the Particle Filter to obtain $\left\{ \boldsymbol{h}_{t}^{\left(i\right)}\right\} _{i=1}^{N}$,
$\left\{ a_{i}\right\} _{i=1}^{N}$, and $\left\{ w_{t}^{\left(i\right)}\right\} _{i=1}^{N}$,
where $w_{t}^{\left(i\right)}$ is the weight of particle $i$ at
time $t$. $a_{i}$ is the ancestor index of particle $i$ at time
$t-1$.
\item Normalised the weights $W_{t}^{\left(i\right)}=\frac{w_{t}^{\left(i\right)}}{\sum w_{t}^{\left(i\right)}}$.
\end{itemize}
\item Update the $m_{t}^{\left(i\right)}$ and $n_{t}^{\left(i\right)}$
as follows
\[
m_{t}^{\left(i\right)}=\lambda m_{t-1}^{\left(k_{i}\right)}+\left(1-\lambda\right)S_{t-1}+\nabla\log g_{\theta}\left(\boldsymbol{y}_{t}|\boldsymbol{h}_{t}^{\left(i\right)}\right)+\nabla\log p_{\theta}\left(\boldsymbol{h}_{t}^{\left(i\right)}|\boldsymbol{h}_{t-1}^{\left(k_{i}\right)}\right)
\]

\item Update the score vector
\[
S_{t}=\sum_{i=1}^{N}W_{t}^{\left(i\right)}m_{t}^{\left(i\right)}
\]
\end{itemize}
\end{algorithm}
Setting $\lambda=1$ gives the \citet{Poyiadjis:2011} algorithm.
\citet{Nemeth:2016} shows that bias and variance of both score estimate
vary according to $\lambda$. Reducing the value of $\lambda$ has
the effect of increasing the bias, but it reduces the Monte Carlo
variance of estimates. They also show that by setting $\lambda\approx0.95$
will produce an estimate for the score with linearly increasing variance
and minimal bias. We use $\lambda=0.95$ in all our application.

\section{Sampling the scaling parameters in the deep interweaving representations\label{sec:deep interweaving}}

In the deep interweaving representation, we sample the scaling parameter
$B_{jj}$ indirectly through $\mu_{2j}$, $j=1,...,k$. The implied
prior $p\left(\mu_{2j}\right)\propto\exp\left(\mu_{2j}/2-\exp\left(\mu_{2j}\right)/2\right)$
and the density $p\left(B_{.,j}^{*}|\mu_{2j}\right)\sim N_{k_{j}}\left(0,\exp\left(-\mu_{2j}\right)I_{k_{j}}\right)$
and the likelihood given by Equation \eqref{eq:likelihood SV2} yields
the posterior
\[
p\left(\mu_{2j}|B_{.,j}^{*},h_{2j,.}^{*},\phi_{2j},\tau_{2j}^{2}\right)\propto p\left(h_{2j,.}^{*}|\mu_{2j},\phi_{2j},\tau_{2j}^{2}\right)p\left(B_{.,j}^{*}|\mu_{2j}\right)p\left(\mu_{2j}\right),
\]
which is not in recognisable form. As in \citet{Kastner:2017}, we
draw a proposal for $\mu_{2j}^{prop}$ from $N\left(A,B\right)$ where
\[
A=\frac{\sum_{t=2}^{T-1}h_{2j,t}^{*}+\left(h_{2j,T}-\phi_{2j}h_{2j,1}\right)/\left(1-\phi_{2j}\right)}{T+1/B_{0}},B=\frac{\tau_{2j}^{2}/\left(1-\phi_{2j}\right)^{2}}{T+1/B_{0}}.
\]
Denoting the current value $\mu_{2j}$ by $\mu_{2j}^{old}$, the new
value $\mu_{2j}^{prop}$ gets accepted with probability $\min\left(1,R\right)$,
where
\[
R=\frac{p\left(\mu_{2j}^{prop}\right)p\left(h_{2j,1}^{*}|\mu_{2j}^{prop},\phi_{2j},\tau_{2j}^{2}\right)p\left(B_{.,j}^{*}|\mu_{2j}^{prop}\right)}{p\left(\mu_{2j}^{old}\right)\left(h_{2j,1}^{*}|\mu_{2j}^{old},\phi_{2j},\tau_{2j}^{2}\right)p\left(B_{.,j}^{*}|\mu_{2j}^{old}\right)}\times\frac{p_{aux}\left(\mu_{2j}^{old}|\phi_{2j},\tau_{2j}^{2}\right)}{p_{aux}\left(\mu_{2j}^{prop}|\phi_{2j},\tau_{2j}^{2}\right)},
\]
where
\[
p_{aux}\left(\mu_{2j}^{old}|\phi_{2j},\tau_{2j}^{2}\right)\sim N\left(0,B_{0}\tau_{2j}^{2}/\left(1-\phi_{2j}\right)^{2}\right).
\]
When normal-gamma prior is used, the density of $B_{.,j}^{*}|\mu_{2j}$
is given by
\[
p\left(B_{.,j}^{*}|\mu_{2j}\right)\sim N_{k_{j}}\left(0,\Psi_{.j}\exp\left(-\mu_{2j}\right)\right),
\]
where $\Psi_{.j}=diag\left(\sigma_{1j}^{2},...,\sigma_{pj}^{2}\right)$.

\section{Simulation Results\label{sec:Simulation-Results}}

\begin{table}[H]
\caption{Multivariate Factor SV model with leverage Simulation Results: Mixed
Sampler with $N=500$ \label{tab:Multivariate-Factor-SV sim result mixed}}

\centering{}%
\begin{tabular}{cccccccccccc}
\hline
 & IACT &  & IACT &  & IACT &  & IACT &  & IACT &  & IACT\tabularnewline
\hline
$\phi_{1}$ & 12.16 & $\tau_{1}^{2}$ & 17.92  & $\mu_{1}$ & 1.89  & $\rho_{1}$ & 12.31  & $\beta_{11}$ & 6.33 & $\beta_{22}$ & 5.86 \tabularnewline
$\phi_{2}$ & 8.85  & $\tau_{2}^{2}$ & 16.43  & $\mu_{2}$ & 1.14  & $\rho_{2}$ & 20.25  & $\beta_{21}$ & 6.32 & $\beta_{23}$ & 5.41 \tabularnewline
$\phi_{3}$ & 11.51  & $\tau_{3}^{2}$ & 14.59  & $\mu_{3}$ & 1.36  & $\rho_{3}$ & 23.23  & $\beta_{31}$ & 6.46  & $\beta_{24}$ & 6.24\tabularnewline
$\phi_{4}$ & 5.33 & $\tau_{4}^{2}$ & 15.88 & $\mu_{4}$ & 1.05 & $\rho_{4}$ & 15.85 & $\beta_{41}$ & 6.37 & $\beta_{25}$ & 5.13 \tabularnewline
$\phi_{5}$ & 4.70  & $\tau_{5}^{2}$ & 8.83  & $\mu_{5}$ & 1.02  & $\rho_{5}$ & 30.09  & $\beta_{51}$ & 6.44  & $\beta_{26}$ & 5.12 \tabularnewline
$\phi_{6}$ & 14.37  & $\tau_{6}^{2}$ & 19.56  & $\mu_{6}$ & 1.31  & $\rho_{6}$ & 24.93  & $\beta_{61}$ & 6.30  & $\beta_{27}$ & 5.52 \tabularnewline
$\phi_{7}$ & 6.78  & $\tau_{7}^{2}$ & 13.70  & $\mu_{7}$ & 1.10  & $\rho_{7}$ & 37.98  & $\beta_{71}$ & 6.32  & $\beta_{28}$ & 5.84\tabularnewline
$\phi_{8}$ & 4.66 & $\tau_{8}^{2}$ & 9.25 & $\mu_{8}$ & 1.04 & $\rho_{8}$ & 30.94 & $\beta_{81}$ & 6.22 & $\beta_{29}$ & 6.11 \tabularnewline
$\phi_{9}$ & 6.17  & $\tau_{9}^{2}$ & 13.30  & $\mu_{9}$ & 1.04  & $\rho_{9}$ & 30.43  & $\beta_{91}$ & 5.72  & $\beta_{2,10}$ & 6.27\tabularnewline
$\phi_{10}$ & 8.12  & $\tau_{10}^{2}$ & 16.65  & $\mu_{10}$ & 1.45  & $\rho_{10}$ & 19.90  & $\beta_{10,1}$ & 6.26 &  & \tabularnewline
$\phi_{f1}$ & 10.36  & $\tau_{f1}^{2}$ & 12.64  &  &  &  &  &  &  &  & \tabularnewline
$\phi_{f2}$ & 13.96 & $\tau_{f2}^{2}$ & 18.24 &  &  &  &  &  &  &  & \tabularnewline
\hline
\end{tabular}
\end{table}

\begin{table}[H]
\caption{Multivariate Factor SV model with leverage Simulation Results: PG
Sampler with $N=500$ \label{tab:Multivariate-Factor-SV sim result PG}}

\centering{}%
\begin{tabular}{cccccccccccc}
\hline
 & IACT &  & IACT &  & IACT &  & IACT &  & IACT &  & IACT\tabularnewline
\hline
$\phi_{1}$ & 87.72  & $\tau_{1}^{2}$ & 229.96  & $\mu_{1}$ & 3.07  & $\rho_{1}$ & 17.16  & $\beta_{11}$ & 5.31  & $\beta_{22}$ & 7.17 \tabularnewline
$\phi_{2}$ & 128.27  & $\tau_{2}^{2}$ & 273.72  & $\mu_{2}$ & 1.04  & $\rho_{2}$ & 23.46  & $\beta_{21}$ & 5.47  & $\beta_{23}$ & 6.71 \tabularnewline
$\phi_{3}$ & 96.86  & $\tau_{3}^{2}$ & 188.35  & $\mu_{3}$ & 1.64  & $\rho_{3}$ & 32.85  & $\beta_{31}$ & 5.43  & $\beta_{24}$ & 6.05\tabularnewline
$\phi_{4}$ & 83.42 & $\tau_{4}^{2}$ & 336.61 & $\mu_{4}$ & 1.02 & $\rho_{4}$ & 24.44 & $\beta_{41}$ & 5.39 & $\beta_{25}$ & 4.94 \tabularnewline
$\phi_{5}$ & 30.09  & $\tau_{5}^{2}$ & 214.09  & $\mu_{5}$ & 1.08  & $\rho_{5}$ & 40.87  & $\beta_{51}$ & 5.39  & $\beta_{26}$ & 6.26 \tabularnewline
$\phi_{6}$ & 271.72  & $\tau_{6}^{2}$ & 510.85  & $\mu_{6}$ & 1.23  & $\rho_{6}$ & 32.51  & $\beta_{61}$ & 5.44  & $\beta_{27}$ & 7.24 \tabularnewline
$\phi_{7}$ & 53.51  & $\tau_{7}^{2}$ & 234.28  & $\mu_{7}$ & 1.12  & $\rho_{7}$ & 45.32  & $\beta_{71}$ & 5.45  & $\beta_{28}$ & 7.64\tabularnewline
$\phi_{8}$ & 27.35 & $\tau_{8}^{2}$ & 174.22 & $\mu_{8}$ & 1.02 & $\rho_{8}$ & 30.07 & $\beta_{81}$ & 5.48 & $\beta_{29}$ & 7.52 \tabularnewline
$\phi_{9}$ & 49.86  & $\tau_{9}^{2}$ & 188.67  & $\mu_{9}$ & 1.02  & $\rho_{9}$ & 42.76  & $\beta_{91}$ & 5.28  & $\beta_{2,10}$ & 7.92\tabularnewline
$\phi_{10}$ & 95.05  & $\tau_{10}^{2}$ & 259.60  & $\mu_{10}$ & 1.17  & $\rho_{10}$ & 19.33  & $\beta_{10,1}$ & 6.34 &  & \tabularnewline
$\phi_{f1}$ & 42.32 & $\tau_{f1}^{2}$ & 121.84  &  &  &  &  &  &  &  & \tabularnewline
$\phi_{f2}$ & 99.06 & $\tau_{f2}^{2}$ & 202.43 &  &  &  &  &  &  &  & \tabularnewline
\hline
\end{tabular}
\end{table}

\begin{table}[H]
\caption{Multivariate Factor SV model with leverage Simulation Results: PGAS
Sampler with $N=500$\label{tab:Multivariate-Factor-SV sim result PGAS}}

\centering{}%
\begin{tabular}{cccccccccccc}
\hline
 & IACT &  & IACT &  & IACT &  & IACT &  & IACT &  & IACT\tabularnewline
\hline
$\phi_{1}$ & 44.55  & $\tau_{1}^{2}$ & 100.13 & $\mu_{1}$ & 1.26  & $\rho_{1}$ & 9.56  & $\beta_{11}$ & 5.53 & $\beta_{22}$ & 7.20 \tabularnewline
$\phi_{2}$ & 22.34  & $\tau_{2}^{2}$ & 81.51  & $\mu_{2}$ & 1.18  & $\rho_{2}$ & 14.33  & $\beta_{21}$ & 5.54  & $\beta_{23}$ & 4.81 \tabularnewline
$\phi_{3}$ & 57.44  & $\tau_{3}^{2}$ & 100.61  & $\mu_{3}$ & 1.36  & $\rho_{3}$ & 18.13  & $\beta_{31}$ & 5.67  & $\beta_{24}$ & 5.79\tabularnewline
$\phi_{4}$ & 34.07 & $\tau_{4}^{2}$ & 126.11 & $\mu_{4}$ & 1.02 & $\rho_{4}$ & 11.96 & $\beta_{41}$ & 5.56 & $\beta_{25}$ & 5.30 \tabularnewline
$\phi_{5}$ & 24.27  & $\tau_{5}^{2}$ & 72.48  & $\mu_{5}$ & 1.11  & $\rho_{5}$ & 24.22  & $\beta_{51}$ & 5.61  & $\beta_{26}$ & 6.92 \tabularnewline
$\phi_{6}$ & 49.25  & $\tau_{6}^{2}$ & 103.15  & $\mu_{6}$ & 1.22  & $\rho_{6}$ & 21.72  & $\beta_{61}$ & 5.50  & $\beta_{27}$ & 7.49 \tabularnewline
$\phi_{7}$ & 20.92  & $\tau_{7}^{2}$ & 61.51  & $\mu_{7}$ & 1.04 & $\rho_{7}$ & 20.34  & $\beta_{71}$ & 5.53  & $\beta_{28}$ & 7.94\tabularnewline
$\phi_{8}$ & 17.76 & $\tau_{8}^{2}$ & 83.77 & $\mu_{8}$ & 1.04 & $\rho_{8}$ & 22.47 & $\beta_{81}$ & 5.46 & $\beta_{29}$ & 8.58 \tabularnewline
$\phi_{9}$ & 24.50  & $\tau_{9}^{2}$ & 72.55  & $\mu_{9}$ & 1.04  & $\rho_{9}$ & 17.18  & $\beta_{91}$ & 5.02  & $\beta_{2,10}$ & 10.17\tabularnewline
$\phi_{10}$ & 69.57  & $\tau_{10}^{2}$ & 184.45  & $\mu_{10}$ & 1.19  & $\rho_{10}$ & 12.84  & $\beta_{10,1}$ & 5.74  &  & \tabularnewline
$\phi_{f1}$ & 21.62  & $\tau_{f1}^{2}$ & 68.15  &  &  &  &  &  &  &  & \tabularnewline
$\phi_{f2}$ & 65.94 & $\tau_{f2}^{2}$ & 146.21 &  &  &  &  &  &  &  & \tabularnewline
\hline
\end{tabular}
\end{table}

\bibliographystyle{apalike}
\bibliography{references_v1}

\end{document}